\documentclass[a4paper,fleqn,usenatbib]{mnras}

\usepackage{mathptmx}

\usepackage[T1]{fontenc}
\usepackage{ae,aecompl}
 \usepackage{etex}
 \usepackage{caption}

\usepackage{graphicx}	
\usepackage{amsmath}	
\usepackage{amssymb}	
\usepackage[utf8]{inputenc} 

\usepackage[normalem]{ulem}
\usepackage{color}
\definecolor{pink}{rgb}{0.98, 0.38, 0.5}

\definecolor{mygreen}{rgb}{0,0.6,0}

\title[Cold dark energy from clusters]{\LARGE Cold dark energy constraints from the abundance of galaxy clusters}

\author[C. Heneka et al.]{Caroline Heneka,$^{1,2,3}$\thanks{E-mail: heneka@thphys.uni-heidelberg.de} David Rapetti,$^{1,4,5,6,7}$  Matteo Cataneo,$^{1,8}$  Adam B. Mantz,$^{9,10}$ 
 \newauthor  Steven W. Allen$^{9,10,11}$ and Anja von der Linden$^{1,9,10,12}$  \vspace{0.25cm} \\
$^{1}$Dark Cosmology Center, Niels Bohr Institute, University of Copenhagen, Juliane Maries Vej 30, DK-2100 Copenhagen, Denmark \\
$^{2}$Department of Physics $\&$ Astronomy, University of California, Irvine, CA 92697, USA \\
$^{3}$Institut f\"ur Theoretische Physik, Ruprecht-Karls-Universit\"at Heidelberg, Philosophenweg 16, D-69120 Heidelberg, Germany \\
$^{4}$Faculty of Physics, Ludwig-Maximilians-Universit\"at, Scheinerstr. 1, D-81679 Munich, Germany  \\
$^{5}$Excellence Cluster Universe, Boltzmannstr. 2, D-85748 Garching, Germany \\
$^{6}$Center for Astrophysics and Space Astronomy, Department of Astrophysical and Planetary Science, \\ \hspace{0.15cm}University of Colorado, Boulder, CO 80309, USA \\
  $^{7}$NASA Ames Research Center, Moffett Field, CA 94035, USA \\
  $^{8}$Institute for Astronomy, University of Edinburgh, Royal Observatory, Blackford Hill, Edinburgh, EH9 3HJ, UK \\
 $^{9}$Kavli Institute for Particle Astrophysics and Cosmology, Stanford University, 452 Lomita Mall, Stanford, CA 94305, USA \\
 $^{10}$Department of Physics, Stanford University, 382 Via Pueblo Mall, Stanford, CA 94305, USA \\
 $^{11}$SLAC National Accelerator Laboratory, 2575 Sand Hill Road, Menlo Park, CA 94025, USA \\
 $^{12}$Department of Physics and Astronomy, Stony Brook University, Stony Brook, NY 11794, USA \\
 }

\date{Accepted 2017 September 28. Received 2017 September 28; in original form 2017 January 25}

\pubyear{2017}

\begin{document}
\label{firstpage}
\pagerange{\pageref{firstpage}--\pageref{lastpage}}
\maketitle

\begin{abstract}
We constrain cold dark energy of negligible sound speed using galaxy cluster abundance observations. In contrast to standard quasi-homogeneous dark energy, negligible sound speed implies clustering of the dark energy fluid at all scales, allowing us to measure the effects of dark energy perturbations at cluster scales. 
We compare those models and set the stage for using non-linear information from semi-analytical modelling in cluster growth data analyses. For this, we recalibrate the halo mass function with non-linear characteristic quantities, the spherical collapse threshold and virial overdensity, that account for model and redshift dependent behaviours, as well as an additional mass contribution for cold dark energy. We present the first constraints from this cold dark matter plus cold dark energy mass function using our cluster abundance likelihood, which self-consistently accounts for selection effects, covariances and systematic uncertainties. We combine cluster growth data with CMB, SNe Ia and BAO data, and find a shift between cold versus quasi-homogeneous dark energy of up to $1\sigma$. We make a Fisher matrix forecast of constraints attainable with cluster growth data from the on-going Dark Energy Survey (DES). For DES, we predict $\sim$50$\%$ tighter constraints on $\left(\Omega_\mathrm{m},w \right)$ for cold dark energy versus $w$CDM models, with the same free parameters.
Overall, we show that cluster abundance analyses are sensitive to cold dark energy, an alternative, viable model that should be routinely investigated alongside the standard dark energy scenario.
\end{abstract}




\section{Introduction} \label{sec:intro}
Cosmology has entered a phase where ever more precise measurements enable us to put increasingly tight constraints on model parameters. Since the first evidence was found for late-time accelerated expansion~\citep{Riess:1998cb,Perlmutter:1998np}, the possibilities of either a cosmological constant, dynamical dark energy or modifications of gravity has been a question central to cosmology \cite[for reviews see e.g.][]{Copeland06,DeFelice:2010aj,Cilfton12,Joyce15}. To guarantee the accuracy of such precise constraints, our modelling toolkit needs to be extended. For this, it is key to explore a wider range of model and parameter spaces, while correctly translating model characteristics into quantities testable against data. The objective is to maximize the information gain and not overlook distinctive signatures.

Both in the linear and non-linear regimes, galaxy cluster surveys are competitive probes of cosmological models and fundamental physics. This has been shown in results ranging from estimating standard cosmological parameters~\citep{2008MNRAS.387.1179M, Vikhlinin09, Mantz10a, 2010ApJ...708..645R, Allen11, 2013ApJ...763..147B, Mantz14, 2015MNRAS.446.2205M, refId0,2016MNRAS.457.1522A,deHaan:2016qvy} to constraining non-Gaussianities of primordial density fluctuations~\citep{2010MNRAS.407.2339S, 2013JCAP...08..004S, 2013MNRAS.434..684M} and testing predictions of General Relativity and modified gravity scenarios~\citep{2009PhRvD..80h3505S, 2010MNRAS.406.1796R, 2012PhRvD..85l4038L, Rapetti12, Cataneo15}. The high-mass end of the halo mass function (HMF), which can be constrained by observations of galaxy clusters, is particularly sensitive to cosmological models through both the background evolution and the linear and non-linear growth of structure. For a vanilla model with a cosmological constant and cold dark matter ($\Lambda$CDM), the HMF has been carefully modelled and calibrated~\citep{ShethTormen99,Tinker08,Jenkins:2000bv,2010ApJ...711..907M,2010ApJ...724..878T,Corasaniti:2010zt,Despali:2015yla,Bocquet:2015pva}. Also, the calibration of mass function parameters with observations has been demonstrated~\citep{Castro:2016jmw,Mehrabi:2016exz}. Continued efforts have been directed to the modelling of the mass function for extended theories~\citep{Schmidt:2008tn,2011ApJ...732..122B,2012MNRAS.424..993C,Barreira:2013xea,Kopp:2013lea,Cataneo:2016iav,Nazari-Pooya:2016bra}. 

For many models beyond $\Lambda$CDM, dark energy can be effectively parametrized as a fluid with an equation of state, $w$, and a sound speed $c_{\mathrm{s}}^2$. When the sound speed is  non-adiabatic, i.e., when dealing with imperfect fluids, $w$ and $c_\mathrm{s}$ are not necessarily directly related to each other as they are in the adiabatic case. For dynamical dark energy models, a sound speed different from one (speed of light), even time- and scale-dependent, is quite natural. One of the simplest examples for dark energy models with a varying sound speed is Quintessence with non-canonical kinetic terms, known as K-essence~\citep{ArmendarizPicon:1999rj,Armendariz00}. Beyond K-essence, in principle any model with an additional propagating scalar degree of freedom exhibits sound speeds different from one, for example the Horndeski class of scalar-tensor theories~\citep{1974IJTP...10..363H,Kobayashi:2011nu,DeFelice:2011hq}.

First attempts to constrain the sound speed of dark energy at the linear level came from~\citet{2003MNRAS.346..987W} using cosmic microwave background (CMB), large-scale structure (LSS) and supernova data, \citet{2004PhRvD..69h3503B} using CMB and CMB LSS cross-correlation data, and~\citet{Hannestad:2005ak} using CMB, galaxy clustering and weak-lensing data. Later,~\cite{Abramo09} forecasted sound speed constraints from galaxy cluster number counts,~\cite{Appleby:2013upa} considered constraints possible from number counts and the CMB, and ~\cite{Hojjati:2015qwa} studied the potential use of CMB lensing data for this purpose. ~\cite{Creminelli:2009mu} and~\cite{Batista:2013oca} extended the analysis of dark energy models with negligible sound speed to the non-linear level of structure formation utilizing the spherical collapse model. ~\citet{2011Basse} and~\citet{Basse:2013zua} followed a similar approach but for dark energy models with an arbitrary speed of sound, and also forecasted parameter constraints.

The goal of this paper is to capture the rich non-linear information of structure formation imprinted on cluster growth data. For this, we implement a semi-analytical framework that incorporates the dominant effects of cold dark energy into the HMF. Using this framework, we present the first observational constraints on cold (clustering) dark energy ($c_{\mathrm{s}}^2=0$). We also compare these results with those for quasi-homogeneous dark energy ($c_{\mathrm{s}}^2=1$) to investigate the impact of the common assumption of $c_{\mathrm{s}}^2=1$ when constraining $w$ and the other relevant standard parameters.

This paper is organized as follows. Section~\ref{sec:cde} describes the semi-analytical framework we use to calculate the non-linear model characteristics of interest for both cold and quasi-homogeneous dark energy, and illustrates their behaviour. In Section~\ref{sec:MF} we proceed to recalibrate the HMF with these non-linear quantities. Section~\ref{sec:data} briefly describes the data we use. In Section~\ref{sec:contour} we present our results on standard cosmological parameters for both cold and quasi-homogeneous dark energy, and in Section~\ref{sec:Fisher} forecast such constraints for DES. We summarize our findings and discuss the broader implications in Section~\ref{sec:end}.

\section{Non-linear characteristics}\label{sec:cde}
In this section we review the effects of assuming cold dark energy on non-linear quantities such as the cosmology-dependent linear threshold of collapse, the virial overdensity and the cold dark energy mass contribution to virialized objects. We use the spherical collapse model to calculate the perturbations stemming from dark energy's being clustering instead of quasi-homogeneous. This enables us to recalibrate the cluster mass function for cold dark energy by incorporating into it these non-linear quantities.

Cold dark energy designates a dark energy fluid whose sound speed is extremely low, i.e. approaching the limit of zero sound speed. A varying, and possibly low, sound speed is natural in scalar-tensor theories, such as for the simple example K-essence, where it avoids pathologies like ghost degrees of freedom or instabilities during the so-called phantom crossing~\citep[a crossing of the $w=-1$ line during the evolution of the K-essence field to negative values;][]{Creminelli:2008wc}. When analysing data, it might therefore be advisable to test models with sound speed zero.

In the following we will assume a dark energy fluid with an effective sound speed $c_\mathrm{s}^2$ and equation of state $w$, at the limits where $c_\mathrm{s}^2 \rightarrow 0$ for cold dark energy and $c_\mathrm{s}^2=1$, i.e., the speed of light, for quasi-homogeneous dark energy. The effective sound speed is defined here as $c_\mathrm{s}^2=\delta p_{\mathrm{e}} / \delta\rho_e$, with $\delta p_{\mathrm{e}}$ and $\delta \rho_e$ being the pressure and density perturbations of the dark energy fluid, respectively.
This relation is more general than that of the adiabatic sound speed $c_{\mathrm{a}}^2$, where $c_{\mathrm{s}}^2 = c_{\mathrm{a}}^2$, with $c_\mathrm{a}^2=w$, only for perfect fluids, but not necessarily for more general theories of dark energy. 
We restrict our current analysis to an effective constant equation of state $w=p_{\mathrm{e}}/ \rho_e$, where $p_{\mathrm{e}}$ is the pressure and $\rho_e$ the energy density of the fluid. Assuming $w$ constant certainly is an approximative choice for models of sound speed different from one; leaving $w$ to vary together with the sound speed will be interesting for future studies. Here we aim at a first analysis exploring the effects of using the two most extreme possibilities for the sound speed. Also, judging from current observational constraints, we expect only a moderate variation of $w$ away from --1.
In addition, we assume negligible anisotropic stress and a flat Friedmann--Lema\^{\i}tre--Robertson--Walker (FLRW) background.

\subsection{Fluid equations and spherical collapse}
Here, we describe the set of equations we employ within the spherical collapse model to derive the non-linear quantities that describe the cosmology-dependent formation of bound structures, which are needed to recalibrate the HMF (Section~\ref{sec:MF}). These are the density threshold of collapse, the overdensity at virialization and a cold dark energy mass contribution to the bound objects. Their behaviour will be discussed in detail in the following Sections~\ref{sec:deltac}--\ref{sec:epsvir}.

The background evolution is governed by the first Friedmann equation in the presence of dark matter and dark energy, $H^2\left ( a\right)=H_0^2 \left[\Omega_{\mathrm{m,0}} a^{-3} + \Omega_{\mathrm{de,0}} e^{-3\left( 1+w\right)}\right]$, with $\Omega_{\mathrm{m,0}}$ and $\Omega_{\mathrm{de,0}}$ being the present-day mean matter and dark energy densities, and $H_0$ the Hubble constant. We neglect the impact of radiation at the late times relevant for this study, and treat $\Omega_\mathrm{m}$ as being made up of CDM + baryons, neglecting baryonic pressure for the collapsing objects. This corresponds to the standard treatment of the fluid equations when calculating collapse threshold and virial overdensity, as they are input for HMFs calibrated with dark-matter-only {\it N}-body simulations.

At late times, all scales relevant for structure formation are well within the horizon and we can safely take the Newtonian limit for the fluid equations. 
The continuity and Euler equation then read
\begin{align}
\frac{\partial \rho_{i}}{\partial t} + \boldsymbol{\nabla}_{\boldsymbol{r}} \cdot \left( \rho_{i} + p_{i} \right)\boldsymbol{v}_i  = 0  \,,
\label{eq:cont} \\
\frac{\partial \boldsymbol{v}_i }{\partial t} +  \left( \boldsymbol{v}_i \cdot \boldsymbol{\nabla}_{\boldsymbol{r}}\right)\boldsymbol{v}_i + \frac{\boldsymbol{\nabla}_{\boldsymbol{r}}p_{i} + \boldsymbol{v}_{i}\dot{p_{i}}}{\rho_{i} + p_{i}} + \boldsymbol{\nabla}_{\boldsymbol{r}}\Phi = 0  \,,
\label{eq:euler}
\end{align}
with density $\rho_{i}$, three-velocity $\boldsymbol{v}_{i}$ and pressure $p_{i}$ for each species $i$, and $\Phi$ denoting the Newtonian potential.

To investigate the non-linear evolution of density fluctuations we expand the fluid quantities, both for dark matter and dark energy \citep{Pace10,Pace14}. Fluctuations in density $\delta_i$ and pressure $\delta p_i$, as well as the peculiar velocity $\boldsymbol{u}_i$, are defined for each species $i$ through $\rho_{i}=\bar{\rho}_i\left( 1+\delta_i \right)$, $p_{i}=\bar{p}_i+\delta p_{i}$, and $\boldsymbol{v}_i = a\left[ H\left( a\right) \boldsymbol{x} + \boldsymbol{u}_i\right]$, respectively, where $\boldsymbol{x}$ is the comoving coordinate and overbars denote the corresponding background quantities. For fluids with constant sound speed $c_{\mathrm{s},i}$ and constant equation of state $w_i$, the corresponding equations for density perturbations and velocity divergence, $\theta_i=\left(\boldsymbol{\nabla}_{\boldsymbol{x}} \cdot \boldsymbol{u}\right)_{i}$, read
\begin{align}
 \dot{\delta}_{i}&+3H\left(c_{\mathrm{s},i}^2-w_i\right)\delta_{i}+\frac{\theta_i}{a}\left[ \left( 1+w_i\right)+\left( 1+c_{\mathrm{s},i}^2\right)\delta_i\right]=\,0  \,,
\label{eq:delta2} \\
 \dot{\theta}_{i}&+2H\theta_{i} + \frac{\theta_{i}^2}{3a}=\,\nabla^2\Phi  \,.
 \label{eq:theta2}
\end{align}
The Poisson equation describes how the potential $\Phi$ is sourced by the density and pressure perturbations, that is
\begin{equation}
\nabla^2\Phi=-4\pi G\sum_I\left( 1+3 c_{\mathrm{s},i}^2\right) a^2\bar{\rho}_{i}\delta_i \,, 
\label{eq:poisson2}
\end{equation}
where the sum runs over each species considered, here dark matter and dark energy. 
In the case of quasi-homogeneous dark energy with $c_\mathrm{s}=1$, the non-linear equations and their linearized counterparts are taken in the limit of negligible dark energy perturbations $\delta_\mathrm{de}\rightarrow 0$. This amounts to neglecting large-scale modes. For cold dark energy with $c_\mathrm{s}\rightarrow 0$ we have negligible dark energy pressure perturbations, $\delta p_{\mathrm{de}} \ll \delta \rho_{\mathrm{de}}$, since $\delta p_{\mathrm{de}}=  c_\mathrm{s}^2 \delta \rho_{\mathrm{de}}$.

For cold dark energy, the combination of the linearized version of equations \eqref{eq:delta2}--\eqref{eq:poisson2} for dark matter and dark energy gives
\begin{align}
\ddot{\delta}_{\mathrm{m}}&+2H \dot{\delta}_{\mathrm{m}} =4\pi G\left( \bar{\rho}_{\mathrm{m}}\delta_{\rm m}+\bar{\rho}_{\rm de} \delta_{\mathrm{de}}\right)   ,\\
\dot{\delta}_{\mathrm{de}}&-3Hw\delta_{\mathrm{de}} =\left( 1+w\right)\dot{\delta}_{\mathrm{m}} \label{eq:delqfori} \,.
\end{align}
Initial conditions are chosen during matter domination when $\delta_{\mathrm{m}}\propto a$ holds. By solving equation~(\ref{eq:delqfori}) in this case, the initial dark energy contrast $\delta_{\mathrm{de},i}$ is obtained in terms of the dark matter density contrast $\delta_{\mathrm{m},i}$ as
\begin{equation}
\delta_{\mathrm{de,}i}=\frac{1+w}{1-3w}\delta_{\mathrm{m,}i} \,, \label{eq:initial}
\end{equation}
with $\dot{\delta}_{\mathrm{m,}i}=H\left( a_{i}\right)\delta_{\mathrm{m,}i}$ at initial scalefactor $a_{i}$, which we fix to $10^{-5}$.
 The initial dark matter density contrast $\delta_{\mathrm{m,}i}$ is adjusted around $\delta_{\mathrm{m},i}\sim10^{-7}$ such that the point of collapse takes place at the low redshifts of interest, as we need to calculate the collapse threshold at the redshifts of our galaxy cluster data. This choice is degenerate with the initial scalefactor $a_i$ chosen. 
 The point of collapse is defined as the singularity where the non-linear matter density perturbation diverges when evolving the set of coupled non-linear equations~(\ref{eq:delta2})--(\ref{eq:poisson2}).
 
Following the non-linear density evolution corresponds to tracking the evolution of a spherical homogeneous tophat overdensity of radius $R$ until its radius reaches zero, i.e. the point of collapse~\citep{GunnGott72}. We also show the radius evolution here, as we employ it to calculate the turnaround radius needed to connect to the radius at virialization through energy conservation (see Section~\ref{sec:delvir}). 
From Birkhoff's theorem, following the radius evolution is equivalent to following the evolution of a separate closed FLRW universe where the scalefactor $a$ is replaced by  a distinct scalefactor $R$ within the overdensity. The radius evolution of the spherical overdensity is obtained from the isotropic and homogeneous solution of the Euler equation~(\ref{eq:euler}) in the Hubble flow, $\boldsymbol{v}=H\boldsymbol{x}$, which corresponds to $\ddot{a}/a = -\boldsymbol{\nabla}\Phi$, with the scalefactor $a$ replaced by the radius $R$ and the gradient potential $\boldsymbol{\nabla}\Phi = \left(4\pi G/3\right)\sum\left( \rho_i + 3p_i\right) \boldsymbol{x}$ inserted. Hence, the evolution of the spherical overdensity in the presence of cold dark energy is described by~\citep{Creminelli10}
\begin{equation}
\frac{\ddot{R}}{R}=-\frac{4\pi G}{3}\left( \rho_{\mathrm{m}}+\rho_{\mathrm{de}}+3\bar{p}_{\mathrm{de}}\right)  \,,
\label{eq:sphR}
\end{equation}
 where we have used $\delta p_{\mathrm{de}} \approx 0$.
For quasi-homogeneous dark energy with $c_\mathrm{s}=1$, we have $\rho_{\mathrm{de}}=\bar{\rho}_{de}$. For the evolution of the dark matter and dark energy densities within a spherical overdensity of radius $R$, the continuity equation~(\ref{eq:cont}) gives 
\begin{equation}
\dot{\rho}_{i}+3\frac{\dot{R}}{R}\left( \rho_{i}+p_{i}\right)=0  \,. 
\label{eq:delQ}
\end{equation}
In order to solve for collapse, equations~(\ref{eq:sphR}) and~(\ref{eq:delQ}) are evolved until a singularity is reached.
 At the initial time $t_i$, we set the radius to $R_{i}=1$, the expansion rate in the linear regime during matter domination to $\left. \mathrm{d}R/\mathrm{d}t \right|_{i}=2\left( 1-\delta_{\mathrm{m},i}/3\right)/(3 t_{i})$, and employ equation~\eqref{eq:initial} for the initial dark energy density contrast. 

We checked that solving for the radius of the spherical overdensity using equations~(\ref{eq:sphR}) and~(\ref{eq:delQ}) gives the same time of collapse as obtained from the non-linear set of equations~(\ref{eq:delta2})--(\ref{eq:poisson2}). In fact, these two approaches are equivalent.
This can for example be shown for dynamical mutation of dark energy, where the effective equation of state within an overdensity or collapsed region is different from the background equation of state~\citep{Abramo07b}. Thus, for the collapse overdensity we have $p_{\mathrm{c}} =  w_{\mathrm{c}} \rho_{\mathrm{c}}$ and $\rho_{\mathrm{c}} = \bar{\rho} \left( 1+\delta\right)$, for which the continuity equation~(\ref{eq:delQ}) reads $\dot{\delta}+ 3\left( 1+\delta\right)\left[ h\left( 1+w_{\mathrm{c}}\right)-H\left( 1+w\right)\right]$, with $h=\dot{R}/R$. This is the same as equation~(\ref{eq:delta2}) when expressing the equation of state inside the overdensity, $w_{\mathrm{c}}$, in terms of the background equation of state via $w_{\mathrm{c}} = w + \left( c_\mathrm{s}^2 - w\right)\delta/\left(1+\delta\right)$.

Note that for dark energy sound speeds different from one or zero, the tophat profile evolution for density perturbations, with which equations~(\ref{eq:delta2})--(\ref{eq:poisson2}) comply, does not hold. The absence of a sharp tophat profile leads to a scale- (or mass-) dependence in the perturbations, which propagates to derived quantities like the density threshold of collapse, so that either an interpolation down to the well-behaved case of sound speed zero or an averaging of the derived quantities, e.g. the threshold of collapse, over a tophat profile are necessary~\citep{2011Basse,Basse:2013zua}. 
This scale- (or mass-) dependence of the perturbations that leads to a breakdown of the tophat approximation can be translated into a Jeans mass, depending on the sound speed of the dark energy fluid. This mass corresponds to a characteristic scale where the effects of dark energy clustering become important. For example, sound speeds of the order of $10^{-4}$ and $10^{-5}$ correspond to masses of the order of $10^{14}M_{\odot}$ and $10^{15}M_{\odot}$, respectively, which are typical masses for galaxy clusters. In this work we opt for comparing cold dark energy of negligible sound speed against the standard quasi-homogeneous dark energy. The advantage is that these two limiting cases are fully consistent with the semi-analytical treatment described above, in which the tophat evolution of the spherical overdensity is physically motivated.

\subsection{Collapse threshold} \label{sec:deltac}
\noindent Here we discuss the cosmology-dependent linear density threshold of collapse for a range of values of the present-day matter density and dark energy equation of state. We include examples with $w$ values differing from -1, that are disfavoured by combined constraints with the CMB but allowed by cluster-only data, to display the qualitative trends of the characteristic quantities of collapsed structures.
 
\begin{figure*}
\centering
\includegraphics[width=0.9\columnwidth]{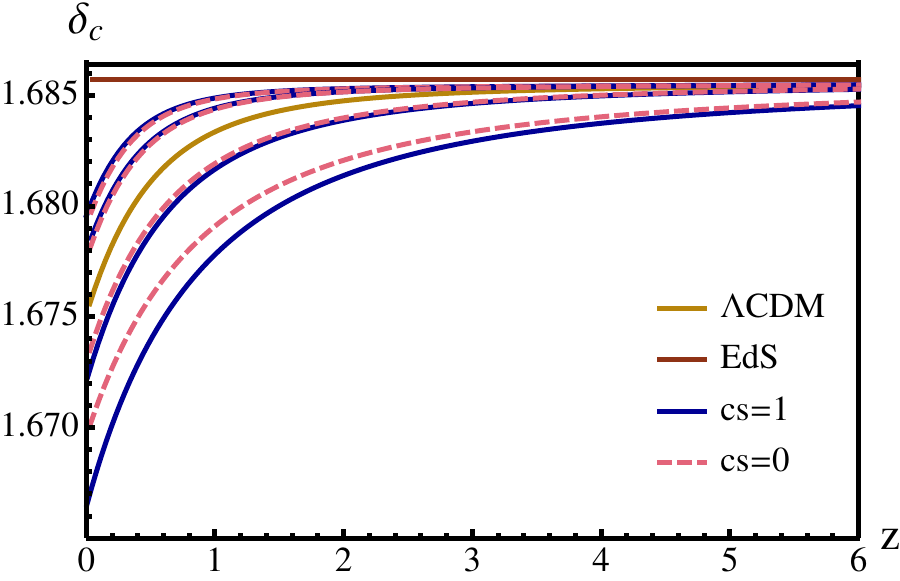}
\hspace{0.2cm}
\includegraphics[width=0.9\columnwidth]{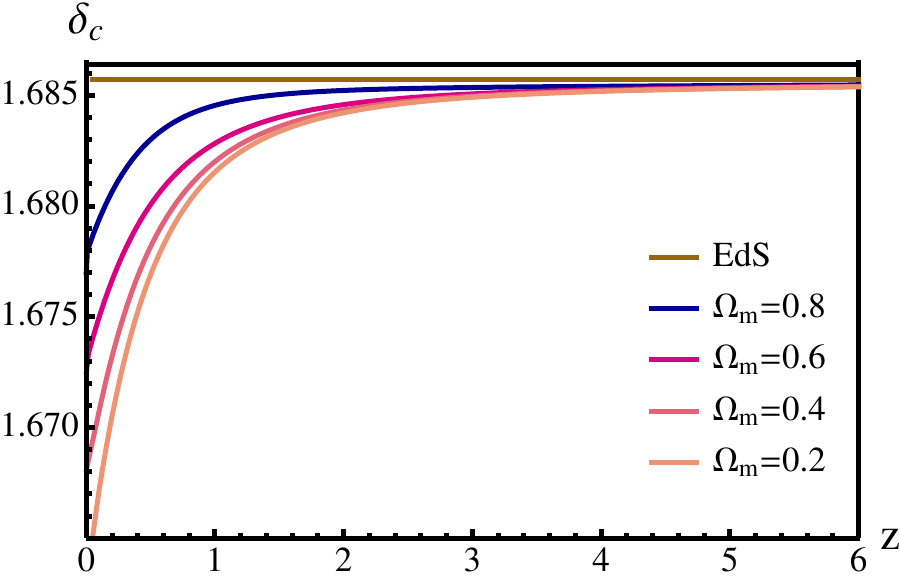}
\vspace{0.25cm}
\\
\hspace{0.01cm}
\includegraphics[width=0.9\columnwidth]{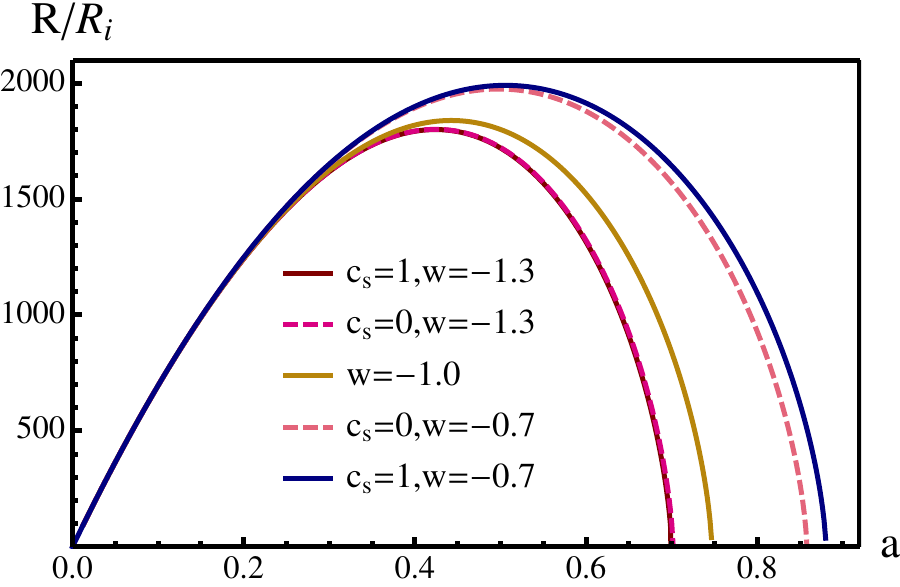}
\hspace{0.2cm}
\includegraphics[width=0.9\columnwidth]{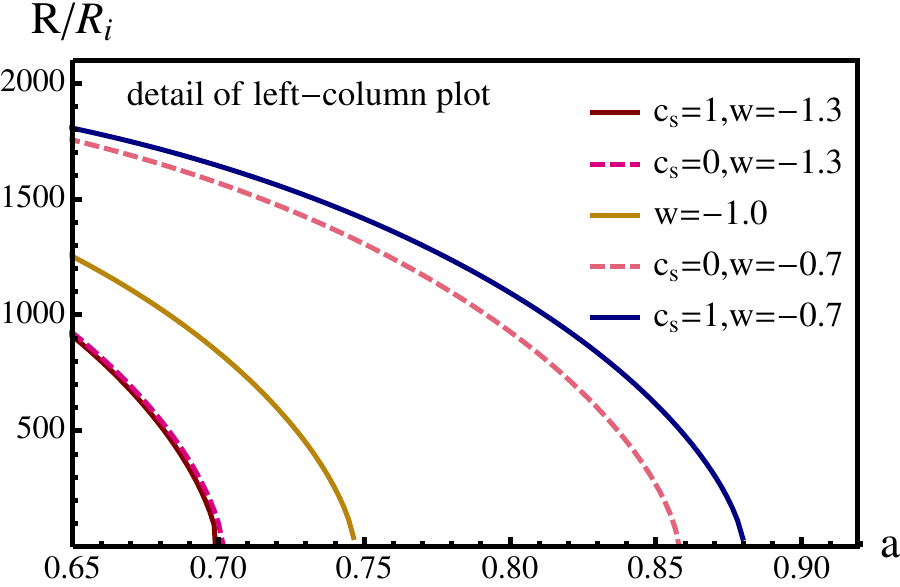}
\vspace{0.25cm}
\\
\hspace{0.01cm}
\includegraphics[width=0.9\columnwidth]{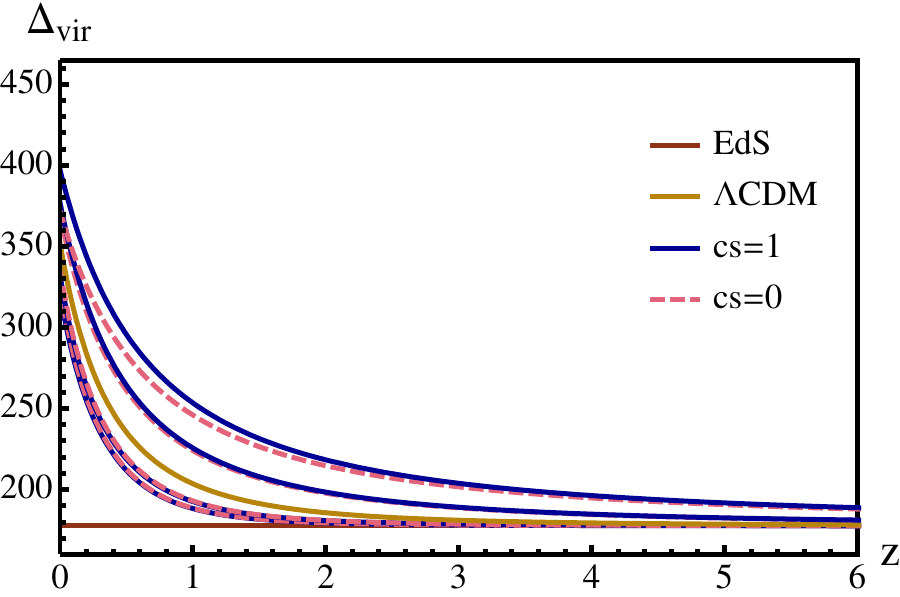}
\hspace{0.2cm}
\includegraphics[width=0.9\columnwidth]{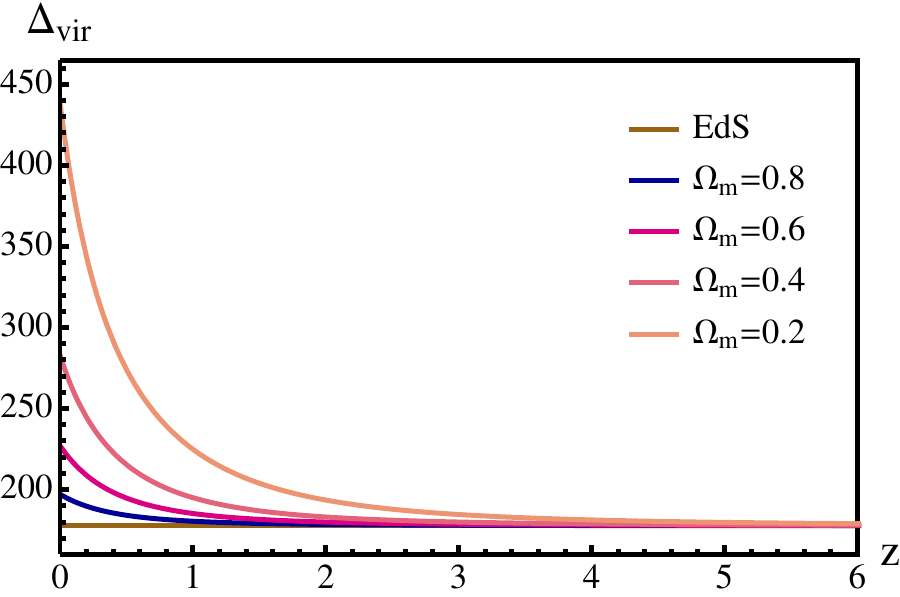}
\caption{\label{FIG:deltac} Top left panel: collapse threshold $\delta_{\mathrm{c}}$ as a function of redshift for $w=-1.4$ to $w=-0.6$ in steps of $0.2$ (top to bottom curves) for fixed $\Omega_\mathrm{m}=0.3$. Solid blue curves correspond to quasi-homogeneous dark energy with sound speed $c_\mathrm{s}=1$, and dashed magenta curves to cold dark energy with sound speed $c_\mathrm{s}=0$. The EdS case with constant $\delta_{\mathrm{c}}=1.686$ is shown in brown. For both $w>-1$ and $w<-1$ the $c_\mathrm{s}=0$ curves are slightly closer to $\Lambda$CDM (middle gold curve) than models with $c_\mathrm{s}=1$. 
Top right panel: $\delta_{\mathrm{c}}\left(z\right)$ for fixed $w=-1$ and varying $\Omega_\mathrm{m}$, as indicated. Increasing $\Omega_\mathrm{m}$ is partly degenerate with shifting from $c_\mathrm{s}=1$ to $c_\mathrm{s}=0$ for $w>-1$ and vice versa for $w<-1$, see the top left panel, which qualitatively corresponds to adding or removing part of the matter contribution. 
Middle left panel:  time evolution of the radius $R$ normalized to the initial radius $R_\mathrm{i}$ of spherical overdensities for $w=-1.3,-1,-0.7$, and fixed $\Omega_\mathrm{m}=0.3$, for $c_\mathrm{s}=1$ (solid curves) and $c_\mathrm{s}=0$ (dashed curves). 
Middle right panel: expanded view of the left-hand panel. Note for $w=-1.3$ the difference between the two sound speeds is considerably smaller than for $w=-0.7$, with the $c_\mathrm{s}=0$ case being only marginally closer to $\Lambda$CDM as compared to $c_\mathrm{s}=1$.
Bottom left panel: virial overdensity $\Delta_{\mathrm{vir}}$ as a function of redshift for $c_\mathrm{s}=1$ (solid) and $c_\mathrm{s}=0$ (dashed), with $w$ varied from $w=-1.4$ to $w=-0.6$ in steps of $0.2$  (top to bottom curves), and fixed $\Omega_\mathrm{m}=0.3$. The EdS value of $\Delta_{\mathrm{vir,EdS}}=18\pi^2$ is shown in brown. Again as for $\delta_\mathrm{c}$, for both $w>-1$ and $w<-1$ the $c_\mathrm{s}=0$ curves are slightly closer to the $\Lambda$CDM case (middle gold curve) as compared to $c_\mathrm{s}=1$. Bottom right panel: $\Delta_{\mathrm{vir}}(z)$ for $\Omega_\mathrm{m}$ as indicated, and $w=-1$ fixed. Note that for $w=-1$ all these quantities are the same for the two speeds of sound, as dark energy perturbations vanish for a cosmological constant in the subhorizon limit. Depicted is the evolution up to $z=6$ to show convergence towards the EdS behaviour at high redshifts where matter dominates the evolution and the spherical collapse model can be solved analytically. For parameter estimation and forecasting in this paper, only redshifts $z < 1$ are relevant.}
\end{figure*}

To solve for the evolution of a spherical overdensity we evolve equations~(\ref{eq:delta2})--(\ref{eq:poisson2}) until the non-linear density perturbation diverges and the point of collapse is reached. The linear density contrast of matter at the time of collapse is the so-called spherical collapse threshold $\delta_{\mathrm{c}}$. In an Einstein-de Sitter (EdS) universe (i.e. $\Omega_{\rm m} =1$), $\delta_{\mathrm{c}}=1.686$, independent of redshift and initial overdensity \citep[see e.g.][]{Peebles80,Bertschinger93}. The inclusion of dark energy modifies the dynamics of the spherical collapse, introducing a redshift-dependence in the threshold of collapse. This redshift dependency of the collapse threshold is displayed in the top left panel of Fig.~\ref{FIG:deltac} for $w \in [-1.4,-0.6]$ (in steps of $0.2$ from top to bottom) for $c_\mathrm{s}^2=1$ and $c_\mathrm{s}^2=0$. Note that for $\Lambda$CDM, i.e. $w=-1$, these two cases coincide, as perturbations vanish for a cosmological constant. For both $c_\mathrm{s}=1$ and $c_\mathrm{s}=0$, $w>-1$ lowers $\delta_{\mathrm{c}}$ with respect to that of $\Lambda$CDM, while dark energy with $w<-1$ has the opposite effect. But, for cold dark energy the collapse threshold is lowered for $w<-1$ and enhanced for $w>-1$ as compared to $\Lambda$CDM, i.e., for both $w>-1$ and $w<-1$ the curves for $c_\mathrm{s}=0$ in comparison to $c_\mathrm{s}=1$ are closer to the $\Lambda$CDM case of $w=-1$. 

As can be seen from the radius evolution of the spherical overdensity in the middle panels of Fig.~\ref{FIG:deltac}, the change of collapse dynamics with $w$ translates into a slightly delayed collapse in a cold dark energy scenario as opposed to quasi-homogeneous dark energy for $w<-1$. This is because dark energy becomes important earlier, hindering the collapse, with $\rho_\mathrm{de} \sim \exp\left[ - H\left( 1+w \right)\right]$ growing for $w<-1$ with a positive argument in the exponential. The opposite is true for $w>-1$, where spherical overdensities collapse earlier, as dark energy starts to dominate later. 
In all cases, cold dark energy tends to bring radius evolution and collapse threshold closer to those of $\Lambda$CDM with respect to quasi-homogeneous dark energy. Also, the modification of the spherical collapse dynamics by the inclusion of dark energy becomes more apparent when further away from an EdS scenario, i.e. the lower $\Omega_\mathrm{m}$ the lower $\delta_{\mathrm{c}}$. Note that an increased $\Omega_\mathrm{m}$ is partly degenerate with shifting from $c_\mathrm{s}=1$ to $c_\mathrm{s}=0$ for $w>-1$ and vice versa for $w<-1$, which is to be expected for clustering dark energy behaving like some kind of extra matter contribution. This behaviour is shown in the top right panel of Fig.~\ref{FIG:deltac} for different values of $\Omega_\mathrm{m}<1$, with $w=-1$ fixed.

\vspace{0.5cm}
\subsection{Virial overdensity} \label{sec:delvir}
In the context of the Press--Schechter formalism~\citep{1974PS} and its extensions (see e.g. \citet{2007Zentner} for a review), a halo mass is defined as the mass enclosed by the virial radius $R_{\rm vir}$, within which there is an interior mean overdensity $\Delta_{\rm vir}$ with respect to a reference background density $\bar\rho_{\rm ref}$. In general, the virial overdensity $\Delta_{\rm vir}$ depends on redshift and cosmology. In contrast, observational masses are usually associated with more convenient fixed overdensities, with typical values such that the interior average density is $\rho_{\rm int} = 300\bar\rho_{\rm m}$ or $\rho_{\rm int} = 500\bar\rho_{\rm crit}$, where $\rho_{\rm crit}=3H_0^2/\left(8\pi G \right)$ is the critical density of the Universe. Thus, to convert between virial and observed masses we need to derive $\Delta_{\rm vir}$ explicitly for each cosmology.

In the spherical collapse model, the redshift of virialization is obtained by relating it to the turnaround of the collapsing sphere via the virial theorem. The time of turnaround is reached when the radius of the sphere is maximal, and equivalently the quantity $\log \left(\delta_{\rm NL}+1 \right)/a^3$, with non-linear density contrast $\delta_{\rm NL}$, is minimal before diverging at the time of collapse. Radius and overdensity at turnaround can then be connected to radius and overdensity at virialization by making use of energy conservation. 
We use the connection between virial and turnaround radius given by~\citet{Lahav91}. 
The virial radius gives the scalefactor of virialization, where we tracked the radius evolution of the sphere as described in Section~\ref{sec:cde}. Having the time and radius of virialization, the virial overdensity $\Delta_{\mathrm{vir}}$ is due to mass conservation given by~\citep{2009PhRvD..80d3530B,2010EdS,2012PhRvD..86j3002M}
\begin{equation}
\Delta_{\mathrm{vir}}=\left( \delta_{\mathrm{NL,vir}} + 1\right)
=\left(\delta_{\mathrm{ta}} +1 \right) \left( \frac{a_{\mathrm{vir}}}{a_{\mathrm{ta}}}\right)^3\left( \frac{R_{\mathrm{ta}}}{R_{\mathrm{vir}}}\right)^3 ,
\end{equation}
where $a_{\mathrm{ta}}$ and $R_{\mathrm{ta}}$ are the turnaround scalefactor and radius, and $a_{\rm vir}$ and $\delta_{\mathrm{NL,vir}}$ are the scalefactor and non-linear density contrast, respectively, at the time of virialization.

The virial overdensity needs to be calculated for every redshift and cosmological parameter set of interest. In order to speed up the calculations, we fit the virial density threshold on a grid of $\Omega_\mathrm{m}$ and $w$, aiming at subpercent accuracy. The fitting formula is an expansion around the EdS case of $\Omega_\mathrm{m}=1$ and constant $\Delta_{\mathrm{vir}}=  18 \pi^2$, with $\Omega_\mathrm{m}\left(z\right)$ order of unity at relevant redshifts. It is normalized for $\Delta_\mathrm{vir}$ to be inversely proportional to $\Omega_\mathrm{m}$. The fitting formula reads
\begin{equation}
\Delta_{\mathrm{vir}}\left( z \right) = \frac{\left[ 18\pi^2 + a\left( 1-\Omega_\mathrm{m}\left( z \right)\right) + b \left( 1-\Omega_\mathrm{m}\left( z\right)\right)^2\right]}{\Omega_\mathrm{m}\left( z \right)} \,,
\end{equation}
with parameters $a$, $b$ and the fractional matter density parameter $\Omega_\mathrm{m}\left( z \right)$. For our HMF calculations in Section~\ref{sec:MF}, we interpolate the values of the virial density threshold on the fitted grid to convert observed cluster masses to virial masses.

The bottom panels of Fig.~\ref{FIG:deltac} show the virial overdensity for different values of $w$ and $\Omega_\mathrm{m}$. As was the case in Section~\ref{sec:deltac} for the collapse threshold, the presence of dark energy tends to increase the virial density threshold compared to the constant EdS case. The effect is bigger the earlier dark energy becomes important.  When comparing cold to quasi-homogeneous dark energy in the bottom left panel of Fig.~\ref{FIG:deltac}, the virial overdensity is larger for $w<-1$, as collapse is hindered, and lower for $w>-1$, with cold dark energy helping the collapse in this case. For a cosmological constant, the bottom right panel of Fig.~\ref{FIG:deltac} makes clear again that lowering $\Omega_\mathrm{m}$ increases the virial overdensity, in accordance with the fitting formula used.

\begin{figure*}[ht!]
\centering 
\includegraphics[width=0.88\columnwidth]{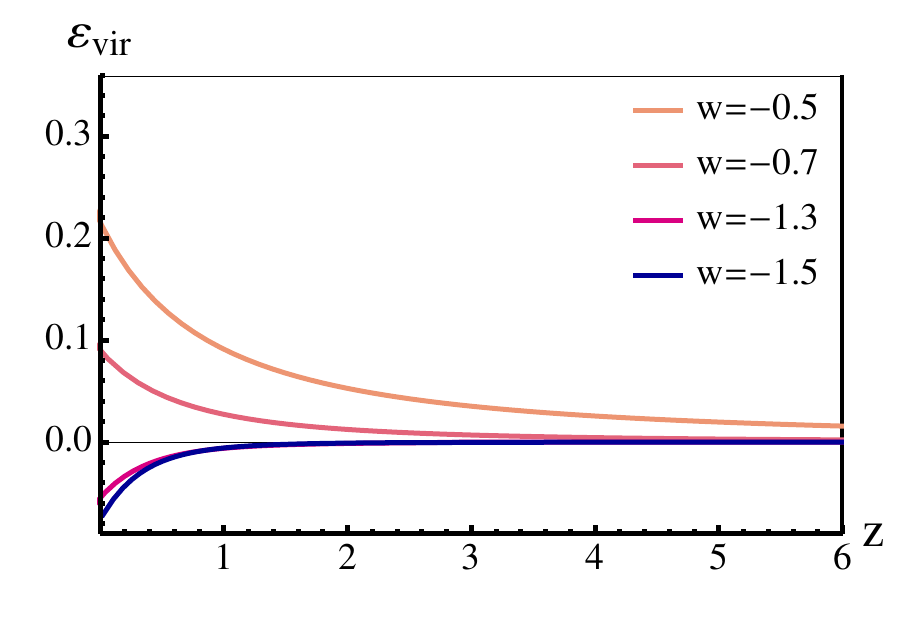}
\hspace{0.2cm}
\includegraphics[width=0.88\columnwidth]{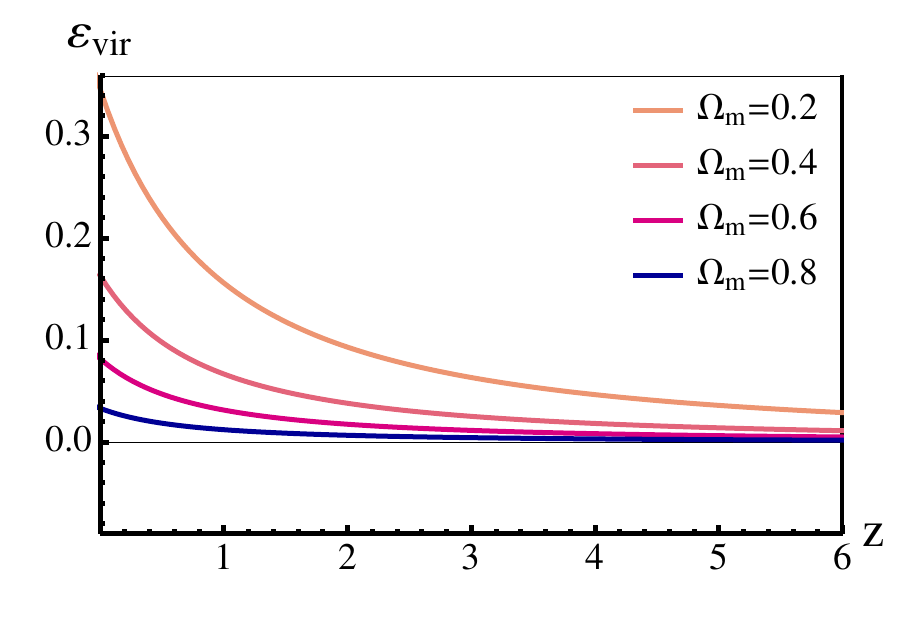}
\caption{Left-hand panel: the ratio $\epsilon_{\mathrm{vir}}\left(z\right)=M_{\mathrm{vir,de}}/M_{\mathrm{vir,m}}$ as a function of redshift. Here $\Omega_{m}$ is fixed to $0.3$ and $w$ is varied as indicated. We can see that the dark energy mass contribution grows as $w$ deviates from $w=-1$ and changes sign from a positive to a negative contribution when going from $w>-1$ to $w<-1$. The negative mass contribution for $w<-1$ points to energy density escaping from dark matter potential wells, so that effectively the total density perturbations become smaller. 
Right-hand panel: $\epsilon_{\mathrm{vir}}(z)$ for $w=-0.5$ and varying $\Omega_{m}$. For these models, the mass correction is higher the lower $\Omega_\mathrm{m}$, as the relative dark energy mass contribution increases; this quantity is larger at lower redshifts for the same reason.}
\label{FIG:epsvir}
\end{figure*}

\subsection{Dark energy mass contribution} \label{sec:epsvir}

In the case of cold dark energy, where dark energy of negligible sound speed is comoving with dark matter, dark energy perturbations can contribute to the total mass of the object. We therefore include the contributions from dark energy perturbations to the total mass by altering the predicted HMF. 
To estimate the extra dark energy contribution to the total mass, we calculate the dark energy mass at virialization as shown in~\citet{Creminelli10}. We assume a tophat profile and calculate the virial radius $R_{\mathrm{vir}}$ as in Section~\ref{sec:delvir}. We define the dark energy mass contribution at virialization as 
\begin{equation}
M_{\mathrm{vir,de}}=4 \pi \int_0^{R_{\mathrm{vir}}} \mathrm{d}R R^2\delta\rho_\mathrm{de} \,, \label{eq:Mvire}
\end{equation}
with dark energy overdensity $\delta\rho_\mathrm{de}$.
The total halo mass $M$ is then rescaled by the dark energy mass contribution as $M\rightarrow M\left(1+\epsilon_{vir,de}\right)$, with the ratio between dark energy and dark matter mass at virialization $\epsilon_{\mathrm{vir}}\left( z \right) = M_{\mathrm{vir,de}}/M_{\mathrm{vir,m}}$, shown in the left-hand panel of Fig.~\ref{FIG:epsvir} for different values of $w$. As expected, the extra dark energy mass contribution grows as $w$ deviates from the $\Lambda$CDM value of $w=-1$, i.e. with increasing inhomogeneity, and changes sign from a positive to a negative contribution when going from the non-phantom $w>-1$ to the phantom $w<-1$ side. The negative mass contribution for $w<-1$ corresponds to dark energy density escaping from dark matter potential wells, so that effectively the total density perturbations are smaller than without dark energy~\citep{2003MNRAS.346..987W}. The right-hand panel of Fig.~\ref{FIG:epsvir} shows the ratio $\epsilon_{\mathrm{vir}}\left( z \right)$ for different values of $\Omega_\mathrm{m}$ and fixed $w=-0.5$. For these models, the positive mass correction is higher the lower $\Omega_\mathrm{m}$, as the dark energy mass contribution increases compared to that of dark matter. Generally, the dark energy contribution also becomes more significant at lower redshifts for the same reason.

An alternative to integrating only over the perturbed dark energy density $\delta_{\rm de}$ is to rescale the total mass by integrating over a dark energy accretion rate for the halo, see again~\citet{Creminelli10}. As a conservative, lower estimate, throughout this paper we will use only the dark energy mass contribution from the dark energy overdensity as evolved at second order in perturbations.

\section{Recalibrated Halo Mass Function} \label{sec:MF}
In this section, we show the impact of cold dark energy on the HMF, i.e., the comoving number density of haloes as a function of mass and redshift, by recalibrating the standard HMF for CDM to account for non-linear perturbative effects stemming from dark energy's being cold (clustering). To do so we include the non-linear characteristic quantities derived in Section~\ref{sec:cde} into our mass function recalibration, as described below. For this, it is crucial to evaluate these quantities, i.e., $\delta_{\mathrm{c}}$, $\Delta_\mathrm{vir}$ and $\epsilon_{\mathrm{vir}}$, for each set of  cosmological parameters when estimating and forecasting constraints in Sections~\ref{sec:contour} and~\ref{sec:Fisher}, respectively.

In practice, we calculate both the Tinker HMF~\citep{Tinker08} and the Sheth--Tormen HMF~\citep{ShethTormen99} for each set of cosmological parameters. We then form the ratio of the Sheth--Tormen HMF of cold dark energy to the Sheth--Tormen HMF of quasi-homogeneous dark energy and multiply it by the standard Tinker HMF. The idea here is that the ratio of Sheth--Tormen HMFs encapsulates the difference between cold and quasi-homogeneous dark energy, i.e., the impact of cold dark energy on the HMF, while the Tinker mass function provides us with a precise estimate for the standard cosmological mass function, informed by {\it N}-body simulations. The ratio of Sheth--Tormen HMFs then includes both effects at the linear level (which is also the case for the Tinker HMF) as well as second-order effects introduced via spherical collapse quantities, which might, due to their scale-dependence, enable us to distinguish the two models.

The recalibrated HMF for the expected number of haloes of mass $M$ at redshift $z$ in a cold dark energy scenario reads
\begin{equation}
\frac{\mathrm{d}n_{\mathrm{cal}}}{\mathrm{d}M}\left( M,z\right)=\frac{\mathrm{d}n_{\mathrm{ST}}/\mathrm{d}M\left(M,z;c_\mathrm{s}=0\right)}{\mathrm{d}n_{\mathrm{ST}}/\mathrm{d}M\left(M,z;c_\mathrm{s}=1\right)}\times\frac{\mathrm{d}n_{\mathrm{T}}}{\mathrm{d}M}\left( M,z\right) \,,
\label{eq:dndmcal}
\end{equation}
with `ST' designating a Sheth--Tormen and `T' a Tinker HMF. 
A similar ratio has been employed in~\citet{Sartoris10}, see also~\citet{Shandera:2013mha} and~\citet{Cataneo15} in order to constrain primordial non-Gaussianity, in the former, 
as first prescribed in~\citet{LoVerde:2007ri}, and to distinguish $f\left(R \right)$ modified gravity theories from GR+$\Lambda$CDM, in the latter.

As a reminder, the functional form of an HMF under spheroidal collapse with mass-independent collapse threshold reads
\begin{equation}
\frac{\mathrm{d}n_T}{\mathrm{d}M}\left( M,z\right)=f\left( \sigma\right)\frac{\bar{\rho}_{\mathrm{m}}}{M}\frac{\mathrm{d}\log{\sigma^{-1}}}{\mathrm{d}M} \,.
\end{equation} 
The Tinker HMF is defined by the multiplicity function $f(\sigma)$~\citep{Tinker08}, which describes non-linear clustering behaviour as a function of matter variance,
\begin{equation} \label{eq:multf}
f\left( \sigma\right)=A \left[ \left( \frac{\sigma}{b}\right)^{-a} +1\right]e^{-c/\sigma^2} \,,
\end{equation}
with the parameters $A$, $a$, $b$ and $c$ fitted using CDM simulations. These parameters depend on the definition of cluster mass as multiples of the overdensity $\Delta$ with respect to $\rho_{\mathrm{crit}}$. The variance $\sigma \left( R\right)$ of the density field smoothed at radius $R$ is defined as
\begin{equation}
\sigma^2\left( R\right) =\frac{1}{2\pi^2}\int{P\left( k\right)\left|W\left( kR\right)\right|^2 k^2 \mathrm{d}k} \,.
\end{equation}
This integrates the matter power spectrum $P\left(k\right)$ over the tophat window function $W\left( kR\right)$ for a sphere of radius $R$ in Fourier space. The theoretical uncertainty of the Tinker HMF from simulations is $\leq 5\%$. We include this uncertainty in our cluster analysis by increasing the width of the Gaussian priors on the parameters of the HMF to $10\%$, as well as accounting for their covariance.

\begin{figure*}
\includegraphics[width=\columnwidth]{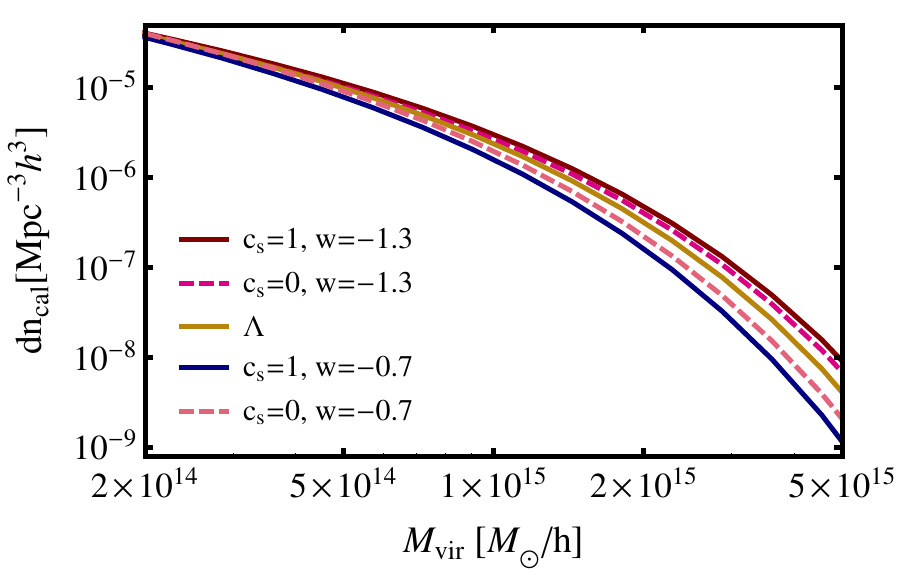}
\hfill
\includegraphics[width=\columnwidth]{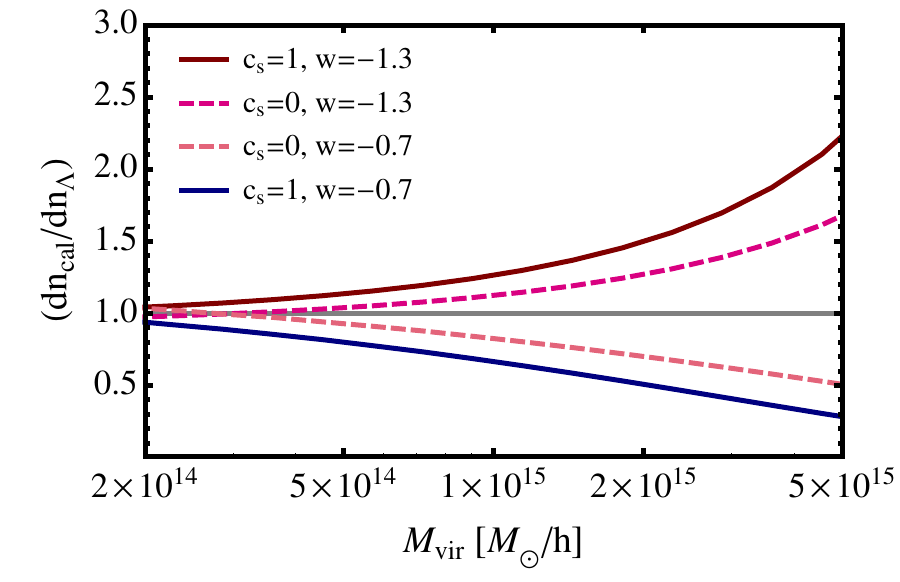}
\hfill
\vspace{0.2cm}
\includegraphics[width=\columnwidth]{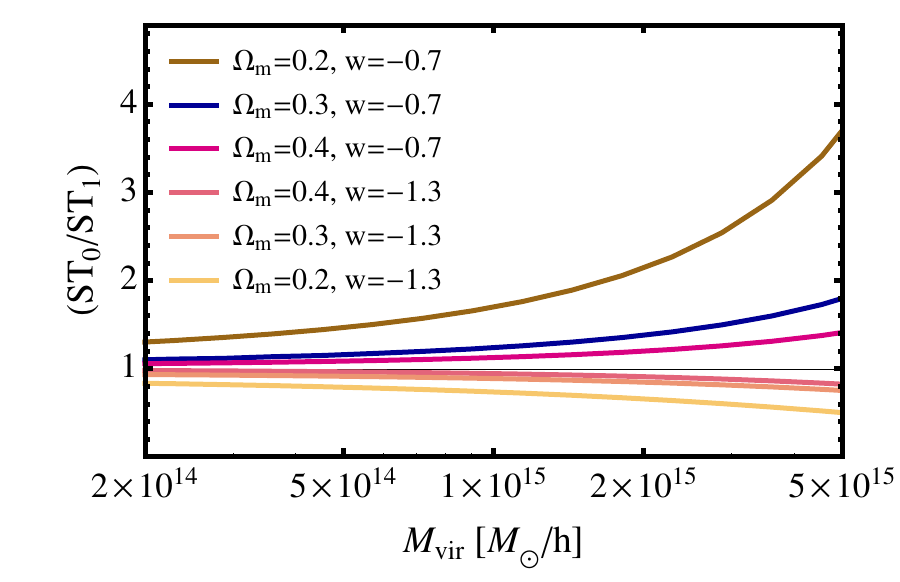}
\hfill
\includegraphics[width=\columnwidth]{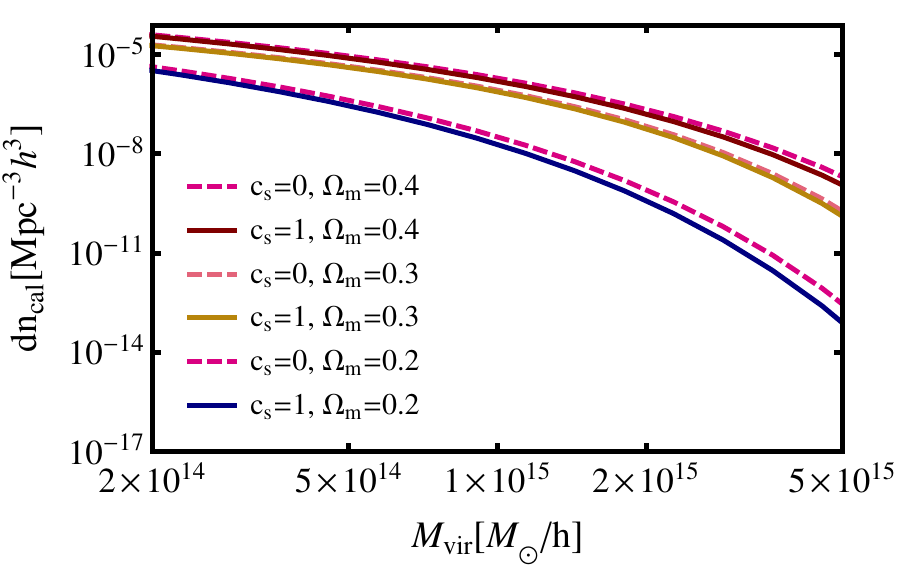}
\caption{Top left panel: recalibrated HMFs at $z=0$ for quasi-homogeneous dark energy ($c_\mathrm{s}=1$; solid lines) and cold dark energy ($c_\mathrm{s}=0$; dashed lines) for $w=-0.7$ (bottom curves) and $w=-1.3$ (top curves), as well as $w=-1$ for which both cases coincide (as dark energy perturbations vanish). $\Omega_\mathrm{m}$ is fixed to $0.3$ for all curves. The HMF is lower for cold  dark energy when $w<-1$, and higher for $w>-1$. 
Top right panel: ratios of recalibrated HMFs with respect to the $\Lambda$CDM case at $z=0$, for $c_\mathrm{s}=0$ (dashed) and $c_\mathrm{s}=1$ (solid). The ratios $>1$ are for $w=-1.3$ and those $<1$ for $w=-0.7$; this is the same in the next panel. The ratios indicate deviations from $\Lambda$CDM by up to a number of times, especially at the high-mass end. 
Bottom left panel: ratios of Sheth--Tormen HMFs for $c_\mathrm{s} =0$ over $c_\mathrm{s} =1$ dark energy at $z=0$ for $\Omega_\mathrm{m}$ values as indicated from the top to the bottom lines. Note the higher and lower number of haloes predicted for cold dark energy for $w<-1$ and $w>-1$, respectively, with the effect growing at high masses and for low $\Omega_\mathrm{m}$. 
Bottom right panel: recalibrated HMFs at $z=0$ for $c_\mathrm{s}=1$ (solid) and $c_\mathrm{s}=0$ (dashed), and for $\Omega_\mathrm{m}=0.2,0.3,0.4$ bottom to top, with $w=-0.7$ fixed. The difference between the sound speeds is more pronounced for low matter densities, as the relative importance of dark energy becomes larger.}
\label{FIG:MFcal}
\end{figure*}

The Sheth--Tormen HMF accounts for ellipsoidal instead of spherical collapse, giving an improved fit to numerical simulations.
The functional form for an HMF under ellipsoidal collapse with mass-dependent peak height $\nu$ states
\begin{equation}
\frac{\mathrm{d}n_{\mathrm{ST}}}{\mathrm{d}M}\left( M,z\right)=\nu f\left( \nu \right)\frac{\bar{\rho}_{\mathrm{m}}}{M^2}\frac{\mathrm{d}\log{\nu}}{\mathrm{d}\log{M}} \,.
\label{eq:ST}
\end{equation}
$f\left(\nu\right)$ depends now on the peak height $\nu = \delta_{\mathrm{c}}/\sigma$, with collapse threshold $\delta_{\mathrm{c}}$, and not solely on $\sigma$ as in the case of spheroidal collapse. As can be seen in equation~(\ref{eq:ST}), this HMF depends on the cosmological background via the mean matter density $\bar{\rho}_m$, as well as on linear density perturbations through $\sigma$ and the linear growth function $D\left(z\right)$. The multiplicity function for the Sheth-Tormen HMF reads~\citep{ShethTormen99}
\begin{equation}
 \nu f\left( \nu \right)=A \sqrt{\frac{2a \nu^2}{\pi}}\left[ 1+ \left( a \nu^2 \right)^{-p}\right]\exp{\left[ -a \nu^2 \right]} \,,
 \end{equation}
with parameters $a$ and $p$ fitted to simulations, where $a$ is determined by the number of massive haloes, $p$ by the shape of the mass function at the low-mass end and $A$ is the normalization ensuring the integral of $f\left(\nu\right)$ over all $\nu$ gives unity. 

When comparing the HMF to observations, the non-linear information on the properties of collapsed structures can be included via $\delta_{\mathrm{c}}$, $\Delta_{\mathrm{vir}}$ and $\epsilon_{\mathrm{vir}}$. 
Generally, these quantities are redshift-dependent, and change with cosmology, and therefore depend on the dark energy sound speed. 
However, when comparing structure formation data to theoretical expectations, the collapse threshold and virial overdensity are sometimes taken at their constant EdS values, or as fitting functions for a $\Lambda$CDM cosmology (see e.g.~\citet{Kitayama:1996ne} and~\citet{Nakamura:1996tk}). 

We go beyond this practice and advocate a more accurate calibration of the model-dependent non-linearities in the HMF by using the results obtained within the spherical collapse model of Sections~\ref{sec:deltac}--\ref{sec:epsvir}. We insert $\delta_{\mathrm{c}}\left( z\right)$ for a given cosmology into the Sheth-Tormen HMF of equation~(\ref{eq:ST}), both for quasi-homogeneous and cold dark energy, and form the ratio in equation~(\ref{eq:dndmcal}). We convert between observed cluster and virial masses, defined using $\Delta_{\mathrm{vir}}$ from Section~\ref{sec:delvir}, assuming a Navarro--Frenk--White (NFW) profile~\citep{NFW96,HuNFW03}. We also include the mass correction at virialization due to dark energy calculated as in Section~\ref{sec:epsvir}. This accounts for differences in virialization, as well as for the mass shift introduced in the HMF due to cold dark energy. 

The standard use of the NFW profile is motivated by {\it N}-body simulations incorporating quasi-homogeneous dark energy in the background, and CDM particles. We expect this profile to also hold reasonably well in the case of cold dark energy, since for negligible sound speed dark energy will comove with dark matter and therefore should not significantly alter the radial profile, while adding more CDM-like substance to the flow. However, this assumption should be checked in detail with future {\it N}-body simulations including cold dark energy.
As we account for here, cold dark energy has an impact on collapsing objects, for example changing the typical collapse overdensities, affecting the number of objects that collapsed in the first place.

In the top left panel of Fig.~\ref{FIG:MFcal}, we show the recalibrated mass function of equation~(\ref{eq:dndmcal}) as a function of virial mass for both $c_\mathrm{s}=1$ and $c_\mathrm{s}=0$, and $w=-1$ ($\Lambda$CDM), $w<-1$ and $w>-1$, with $\Omega_\mathrm{m}=0.3$ fixed for all curves. The predicted cluster abundances are systematically lower for cold compared to quasi-homogeneous dark energy when $w<-1$, and systematically higher for $w>-1$. For $w=-1$ both curves coincide, as dark energy perturbations vanish. This behaviour is to be expected, since the contribution from dark energy perturbations suppresses structure formation for $w<-1$ and enhances it for $w>-1$ with respect to the quasi-homogeneous case. The bottom left panel of Fig.~\ref{FIG:MFcal} underlines these sizeable deviations, showing the ratio of the Sheth--Tormen cluster mass functions of $c_\mathrm{s}=0$ to $c_\mathrm{s}=1$ dark energy for different values of $w$ and $\Omega_\mathrm{m}$. We recalibrate the HMF with this ratio. 

The top right panel of Fig.~\ref{FIG:MFcal} shows the ratio of the recalibrated cluster mass functions for cold and quasi-homogeneous dark energy to the mass function in a $\Lambda$CDM universe for the same set of parameters as in the top left panel. For a cold dark energy scenario, this ratio indicates deviations from $\Lambda$CDM by up to a number of times, depending on parameters chosen. The effect is especially pronounced at the high-mass end where massive clusters of galaxies reside. 

The bottom right panel of Fig.~\ref{FIG:MFcal} shows the dependence on $\Omega_\mathrm{m}$ of the recalibrated cluster mass function, for both $c_\mathrm{s}=1$ and $c_\mathrm{s}=0$ dark energy. We display curves for $\Omega_\mathrm{m}=0.2, 0.3, 0.4$ at fixed $w=-0.7$. As expected, the predicted cluster abundances are systematically higher for cold compared to quasi-homogeneous dark energy as it was in the corresponding case in the bottom left panel for a $w>-1$. Note that the difference between cold and quasi-homogeneous dark energy is more pronounced for low matter densities, as the relative importance of dark energy grows.
 
 Even though here we focused on displaying the parameter dependence of the mass function recalibration on $\Omega_\mathrm{m}$, $w$ and $c_\mathrm{s}$, we stress that the mass function estimate to be obtained with our Markov Chain Monte Carlo (MCMC) data analysis depends on a multitude of parameters, such as those of the mass-observable relations fitted simultaneously with cosmology. Dependencies on astrophysical parameters of course interplay with the ones on cosmological parameters displayed in this section, while our recalibration remains important in order to incorporate the effect of dark energy clustering. In Section~\ref{sec:contour} we will use the recalibrated mass function presented here to constrain cosmological parameters, while marginalizing over astrophysical parameters, with measurements of cluster number counts.

 \section{Data}\label{sec:data}
 For the parameter constraints in Section~\ref{sec:contour} we tested both a cluster-only data set and a combination of cluster, CMB, baryon acoustic oscillation (BAO) and type Ia supernova (SN Ia) data.
The cluster data used here
 consist of three X-ray samples of clusters taken from the ROSAT All-Sky Survey~\citep[RASS;][]{Truemper93}, the Brightest Cluster Sample~\citep[BCS;][]{Ebeling98}, the ROSAT-ESO Flux Limited X-ray sample~\citep[REFLEX;][]{Boehringer:2004fc} and the bright sample of the Massive Cluster Survey~\citep[MACS;][]{Ebeling2010}, together with follow-up X-ray and optical imaging data.\footnote{Note that we do not include measurements of the gas mass fraction $f_{\rm gas}$~\citep{Mantz14}, as the relation between X-ray gas mass, total cluster mass and baryonic fraction has not been investigated yet for cold dark energy cosmology.} 
 To select for high-mass clusters and avoid incompleteness, a flux limit of 0.1--2.4 keV for luminosities $>2.5\times10^{44} h^{-2}$ erg s$^{-1}$ was applied; furthermore systems with X-ray emission dominated by active galactic nuclei have been removed. After cuts, the sample contains 224 clusters. For 94 clusters of the sample, follow-up ROSAT and/or \textit{Chandra} data were used to derive X-ray luminosities, temperatures and gas masses in order to constrain cluster scaling relations.\footnote{X-ray luminosities, temperatures and gas masses are direct measurements of intrinsic quantities that we utilize to fit cluster scaling relations simultaneously with cosmology, as described in detail in~\citet{Mantz10b}.}
 The absolute cluster mass scale is calibrated using weak-lensing measurements for 50 massive clusters taken from the Weighing the Giants program introduced in~\citet{1WtG14},~\citet{2014MNRAS.439...48A} and~\citet{2WtG14}.
 
 When using our cluster data set alone, consisting of RASS catalogues, and follow-up X-ray and lensing data, we also include Gaussian priors on the cosmic baryon density $100\,\Omega_{\mathrm{b}} h^2=2.202\pm0.045$ ~\citep{Cooke14} and the present-day Hubble parameter $h=0.738\pm0.024$~\citep{Riess11}. For the full combination of clusters+CMB+BAO+SNIa, priors on $h$ and $\Omega_{\mathrm{b}}$ are not required. For the CMB data, we combine \textit{Planck}~\citep[1-year release plus {\it WMAP} polarisation data;][]{Planck14c}, together with Atacama Cosmology Telescope~\citep[ACT;][]{Das14} and South Pole Telescope~\citep[SPT;][]{Keisler11,Reichardt12,Story13} measurements at higher multipoles. We also use the Union2.1 compilation of SNIa data~\citep{Union21}, as well as BAO data from the 6-degree Field galaxy Survey~\citep[6dF;][]{Beutler11} and the Sloan Digital Sky Survey~\citep[SDSS release 7 and 9, see][]{Padmanabhan12,Anderson14}. 

In the following, we employ our cluster-only data set and its combination with the above complementary data sets to constrain cosmological parameters for both cold and quasi-homogeneous dark energy, using the recalibrated mass function introduced in the previous section.

\begin{figure*}
\includegraphics[width=0.32\textwidth]{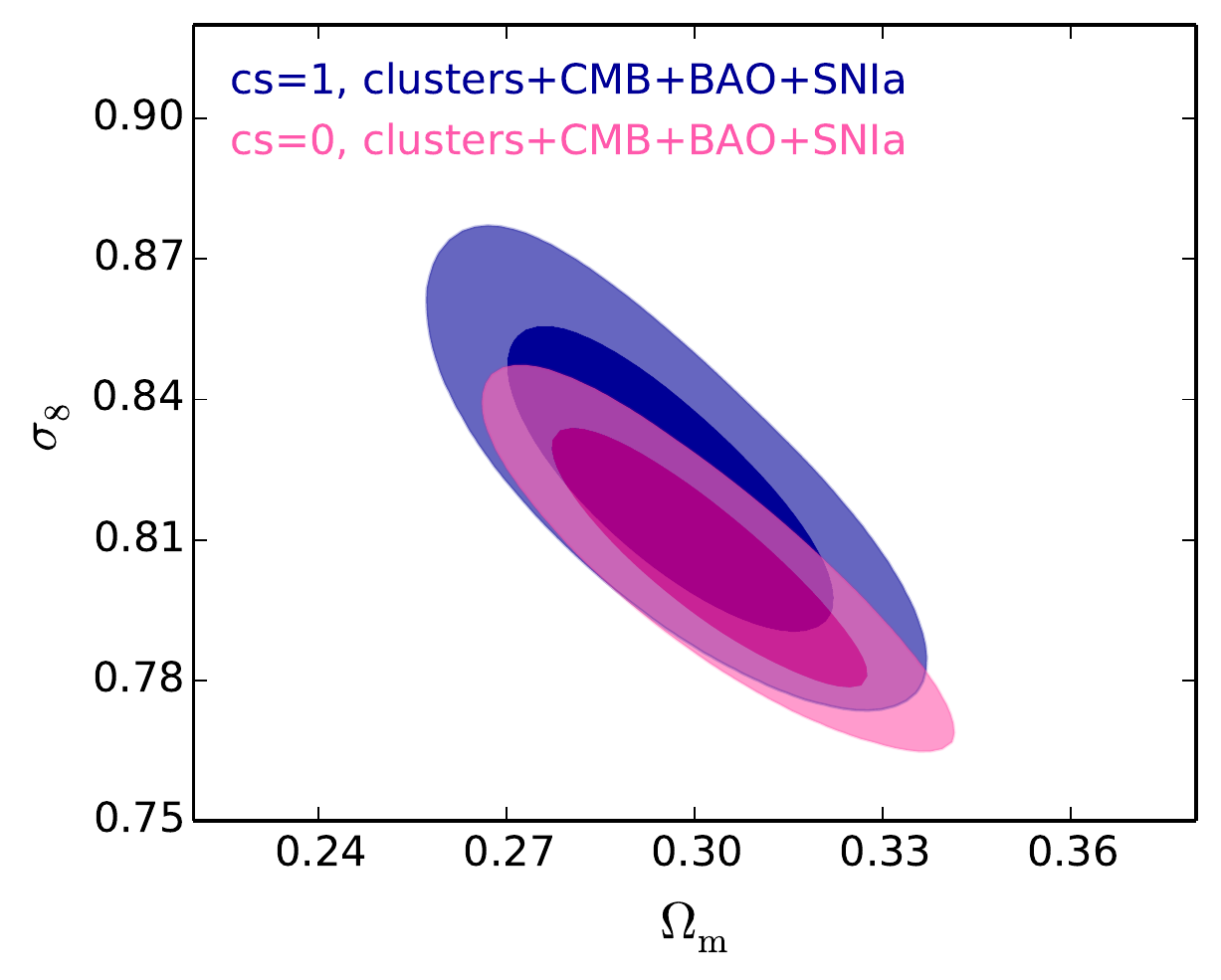}
\label{fig:figure1}
\hspace{0.1cm}
\includegraphics[width=0.32\textwidth]{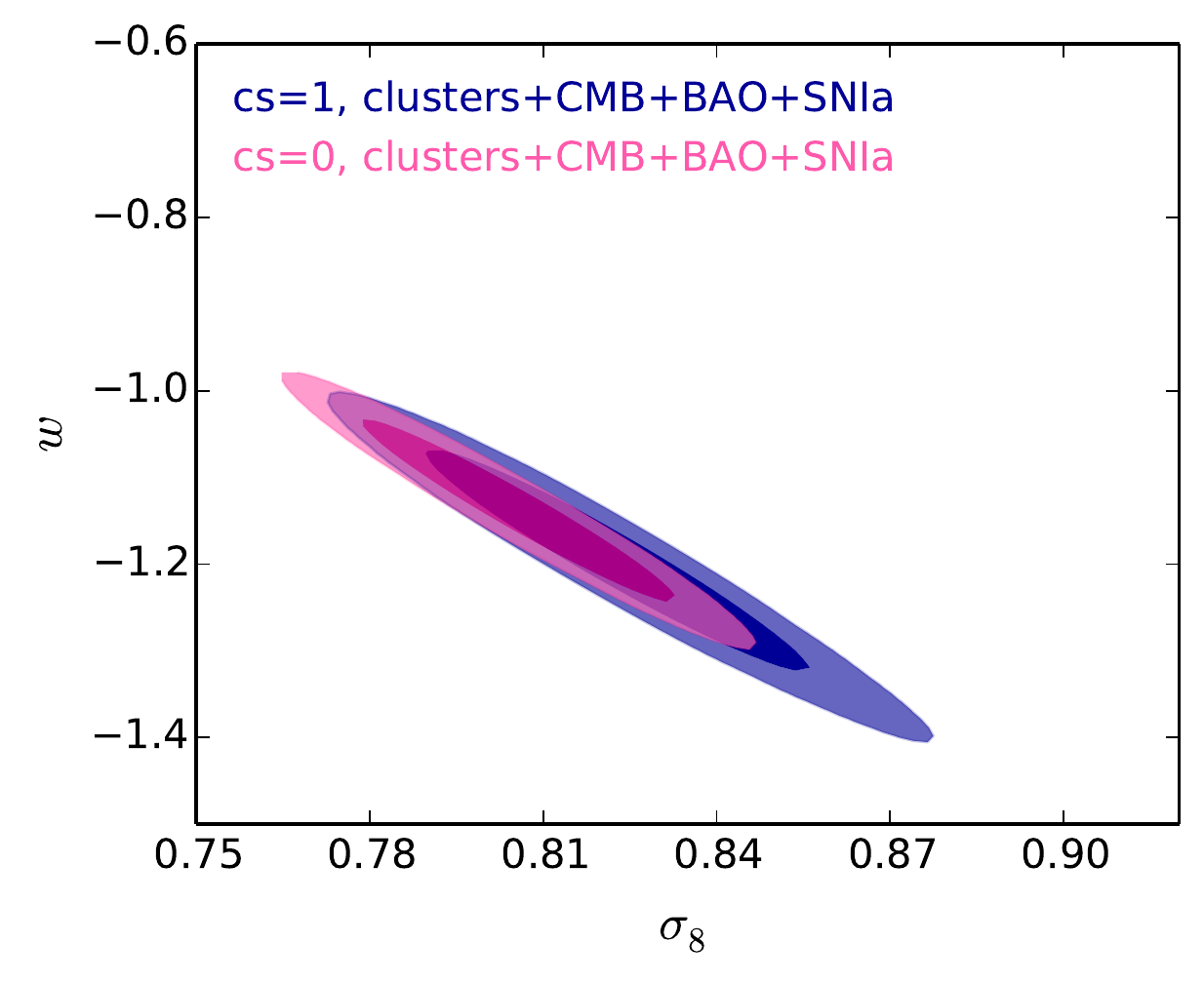}
\label{fig:figure2}
\hspace{0.1cm}
\includegraphics[width=0.32\textwidth]{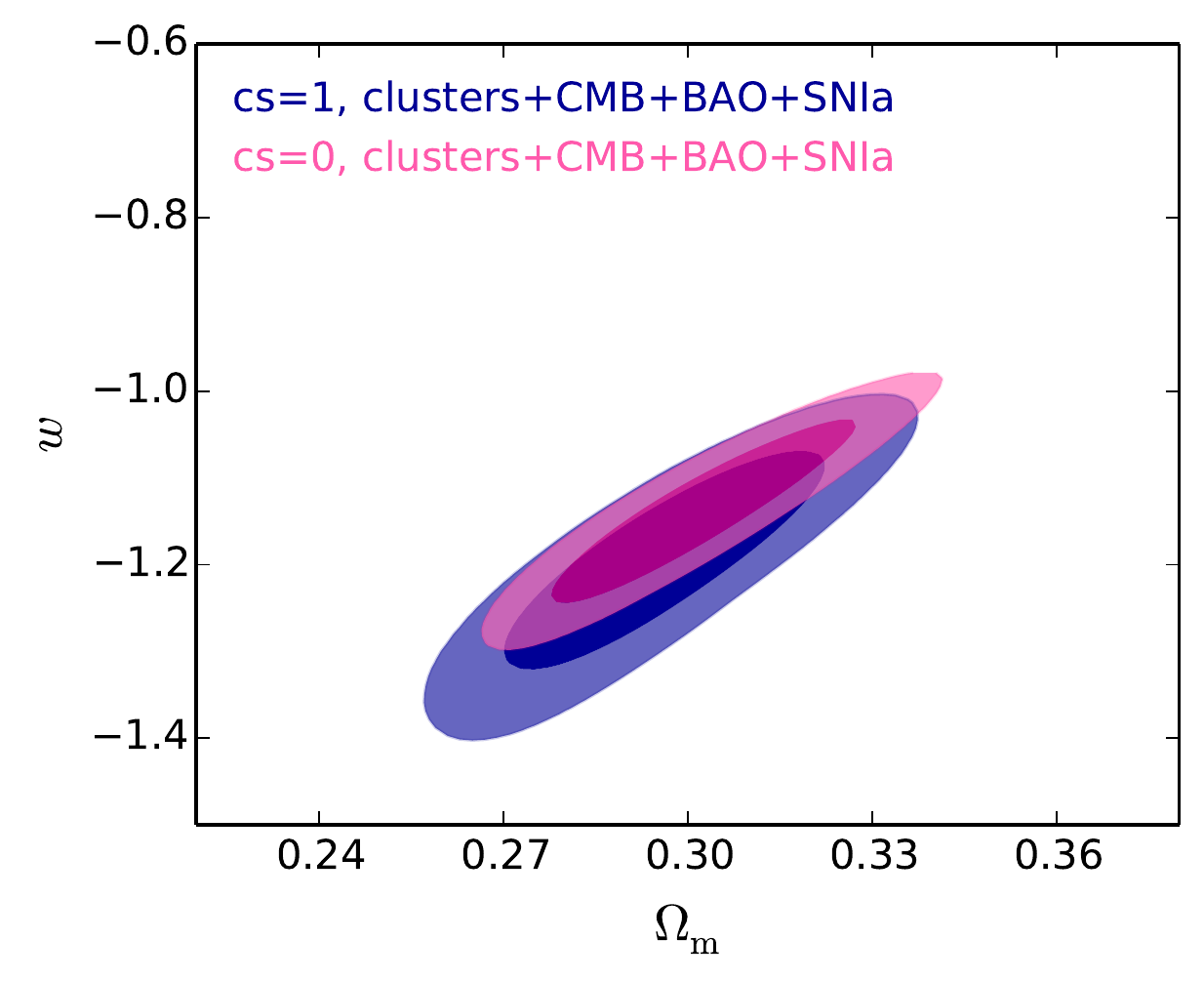}
\caption{Confidence contours for quasi-homogeneous dark energy of sound speed $c_\mathrm{s}=1$ using a Tinker HMF (in blue), and for cold dark energy of sound speed $c_\mathrm{s}=0$ employing a recalibrated mass function (magenta), using a combination of cluster growth data with CMB, BAO and SN Ia data as described in Section~\ref{sec:data}. Dark and light shading indicate the 68.3 and 95.4 per cent confidence regions. A shift for  $c_\mathrm{s}=0$ with respect to  $c_\mathrm{s}=1$ becomes apparent, for example with $w$ shifting towards $\Lambda$CDM, as expected with the HMF for $c_\mathrm{s}=0$ being closer to $\Lambda$CDM (see Fig.~\ref{FIG:MFcal}).}
\label{FIG:contour}
\end{figure*}

\renewcommand{\arraystretch}{1.0}
\setlength{\tabcolsep}{10pt}
    \begin{table*}
    \centering
    \begin{minipage}{0.75\textwidth} 
        \caption{Marginalized best-fitting values and 68.3 per cent confidence intervals for $\sigma_8$, $\Omega_\mathrm{m}$ and $w$, for both cold dark energy with sound speed $c_\mathrm{s}=0$ and quasi-homogeneous dark energy with $c_\mathrm{s}=1$. Results are shown for cluster+CMB+BAO+SNIa data as described in Section~\ref{sec:data}. Note for example the shift to lower $\sigma_8$, higher $\Omega_\mathrm{m}$ and higher $w$ for $c_\mathrm{s}=0$ as compared to $c_\mathrm{s}=1$.}
    \begin{tabular}{ | l |  c| c| c| }
    \hline 
    Analysis & $\sigma_8$ & $\Omega_\mathrm{m} $  & $w$ \\ \hline 
         Clusters + CMB + BAO + SNIa, $c_\mathrm{s}=0$ & 0.806 $\pm$ 0.014 & 0.302 $\pm$ 0.013 & --1.14 $\pm$ 0.05  \\ 
       Clusters + CMB + BAO + SNIa, $c_\mathrm{s} =1$ & 0.823 $\pm$ 0.017 & 0.296 $\pm$ 0.013 & --1.19 $\pm$ 0.07  \\ 
           \hline
    \end{tabular}
        \label{tab:params}
    \end{minipage}
    \end{table*}

\section{Parameter estimation}\label{sec:contour}
Here, we present the first observational constraints on a dark energy model with constant $w$ and $c_{\mathrm{s}}=0$. For this, we employ our cluster growth data and additional complementary data sets, as described in the previous section. Our likelihood analysis and parameter estimation via an MCMC sampler includes theoretical and experimental scatter in the HMF and the mass-observable scaling relations, and also incorporates modified characteristic quantities for structure formation: collapse threshold (Section~\ref{sec:deltac}), virial overdensity (Section~\ref{sec:delvir}) and cold dark energy mass contribution at virialization (Section~\ref{sec:epsvir}). These quantities are interpolated on a grid for computational speed, aiming at per cent accuracy. 

We proceed as follows. For each trial cosmology, the ratio of the Sheth--Tormen HMF for cold versus quasi-homogeneous dark energy (equation~\ref{eq:dndmcal}) is formed, depending both on mass and redshift. In this way, the cosmology-dependent collapse threshold is included into the Sheth--Tormen HMF, on top of the background and linear level calculations taken from {\small \sc CAMB}~\citep{Lewis:1999bs,Howlett:2012mh}.\footnote{Version March 2013; http://camb.info/} 
In addition, the cold dark energy mass contribution is taken into account, resulting in a shift of the mass scale, i.e., adding mass for $w>-1$ and reducing mass for $w<-1$ (see Section~\ref{sec:epsvir}). Equation~(\ref{eq:dndmcal}) is then used as our recalibrated cold dark matter plus cold dark energy mass function to compare with observed cluster number counts. We convert between different mass definitions at differing overdensities assuming an NFW profile~\citep[see appendix in][]{HuNFW03}.
Virial masses are converted to other mass definitions using $\Delta_{\mathrm{vir}}$, as derived in Section~\ref{sec:delvir}, which accounts for differences in virialization between the cold and quasi-homogeneous dark energy models.

We use a modified version of {\small \sc CosmoMC}~\citep{Lewis:2002ah,Lewis:2013hha},\footnote{Version October 2013; http://cosmologist.info/cosmomc/} which incorporates a module that evaluates our cluster growth likelihood~\citep{2015MNRAS.446.2205M} using the data described in Section~\ref{sec:data}. We also employ uniform priors on $\Omega_\mathrm{m} h^2 \in \left[ 0.025,0.3\right]$, $\Omega_{\mathrm{b}} h^2 \in \left[ 0.005,0.1\right]$ and $w \in \left[ -1.5,-0.5\right]$ in order to keep the spherical collapse calculations valid and numerically stable. 
Throughout our analysis, we assume a spatially flat cosmology, the standard effective number of relativistic species $N_{\mathrm{eff}}=3.046$ and the minimal species-summed neutrino mass $\sum m_{\mathrm{\nu}}=0.056$.\footnote{Note that the chosen assumption on the sum of neutrino masses can affect the high-mass end of the cluster mass function. In order to separate out the effect of dark energy clustering, we relied for now on fixing the effective sum of neutrino masses.}

Fig.~\ref{FIG:contour} shows our constraints on $\Omega_\mathrm{m}$, $w$ and $\sigma_8$, the rms variation of the matter density field within 8~$h^{-1}$Mpc spheres, using the combination of cluster growth data with CMB, BAO and SN Ia data, as described in Section~\ref{sec:data}. The blue contours show the results for the standard mass function analysis of quasi-homogeneous dark energy with sound speed $c_\mathrm{s} =1$, employing the Tinker HMF, and the magenta contours those for cold dark energy with sound speed $c_\mathrm{s} =0$. 
For the latter, we employ the recalibrated cluster mass function of Section~\ref{sec:MF} that we implemented into our cluster growth likelihood analysis alongside the standard analysis, which includes simultaneous fitting of cluster scaling relations. For the combination of data sets, a slight shift between the confidence contours for cold versus quasi-homogeneous dark energy is observed, suggesting that this effect will be important to consider with upcoming, more precise measurements. Note this shift puts the $w$ value for $c_\mathrm{s}=0$ closer to the $\Lambda$CDM case, as can be expected from the predicted HMFs relative to $\Lambda$CDM, for example in the top right panel of Fig.~\ref{FIG:MFcal}.

In Table~\ref{tab:params}, we show the corresponding marginalized best-fitting values and 68.3 per cent confidence intervals for $\sigma_8$, $\Omega_\mathrm{m}$ and $w$. We find that for the full combination of cluster+CMB+BAO+SNIa data, differences in the marginalized best-fit values at the 1--2$\,\sigma$-level appear between cold and quasi-homogeneous dark energy. The slight preference for low $w$ values is comparable to that found for other similar data combinations, like {\it Planck}+SNIa+BAO, with best-fitting values of $w\sim-1.1$ for $w$CDM.\footnote{https://wiki.cosmos.esa.int/planckpla/index.php/\newline Cosmological$\_$Parameters (Planck Collaboration XVI 2014)}
Using cluster growth data alone, we find that the derived best-fitting values for cold and quasi-homogeneous dark energy agree within their respective 68.3 per cent confidence intervals, and the estimates obtained for $w$ agree with $\Lambda$CDM at the 1$\sigma$ level.

\section{Fisher forecast} \label{sec:Fisher}
In this section, we use a Fisher matrix formalism to forecast constraints on cosmological parameters from future cluster number counts for the case of both cold and quasi-homogeneous dark energy, in order to test for the impact of our model choice on parameter inference.
For this analysis we use the cosmological parameter vector ${\bf p} = \left\{ \Omega_\mathrm{m},\Omega_\mathrm{b},h,n_\mathrm{s},\sigma_\mathrm{8},w\right\}$, of which we vary only three parameters: $\Omega_\mathrm{m}$, $\sigma_\mathrm{8}$ and $w$. As fiducial values, we take those of the {\it Planck} best fit for the base case $w$CDM lowl+highL+SNLS in \citet{2014A&A...571A..16P},\footnote{PLA/base$\_$w$\_$planck$\_$lowl$\_$lowLike$\_$highL$\_$post$\_$SNLS}$^{,}$\footnote{https://wiki.cosmos.esa.int/planckpla/index.php/File:Grid$\_$limit68.pdf} being $\Omega_\mathrm{m} = 0.28$, $\Omega_\mathrm{b} = 0.044$, $h=0.709$, $n_\mathrm{s} = 0.9581$, $\sigma_\mathrm{8} = 0.87$ and $w=-1.124$. Note also that we do not include a galaxy cluster power spectrum analysis.

The Fisher matrix for cluster number counts, with $N_{l,m}$ number of clusters in the $l$th redshift bin and $m$th observed mass bin, reads~\citep{Holder:2001db}
\begin{equation}
F_{ij} = \sum_{l,m} = \frac{\partial N_{l,m}}{\partial p_{i}} C_{l,m}^{-1}  \frac{\partial N_{l,m}}{\partial p_j} , 
\end{equation}
where the inverse covariance is given by $C_{l,m}^{-1}=1/N_{l,m}$. The $N_{l,m}$ expected for a survey with sky coverage $\Delta \Omega$ is~\citep{Majumdar:2002hd} 
\begin{align}
N_{l,m} = & \frac{\Delta\Omega}{4\pi} \int^{z_{l+1}}_{z_l} \mathrm{d}z \frac{\mathrm{d}V}{\mathrm{d}z} \nonumber \\ & \int^{M^{\mathrm{ob}}_{l,m+1}}_{M^{\mathrm{ob}}_{l,m}} \frac{\mathrm{d}M^{\mathrm{ob}}}{M^{\mathrm{ob}}} \int^{+\infty}_{0} \mathrm{d}M\, n\left( M,z\right) p\left( M^{\mathrm{ob}} | M ;z \right) ,  \label{eq:Nlm}
\end{align}
with comoving volume element per unit redshift interval $\mathrm{d}V/\mathrm{d}z$, HMF $n\left( M,z\right)$ and probability to assign an observed mass $M^{\mathrm{ob}}$ to a cluster of true mass $M$, $p\left( M^{\mathrm{ob}} | M \right)$. The cosmology-dependent comoving volume element is given by 
\begin{equation}
\frac{\mathrm{d}V}{\mathrm{d}z} = \frac{4\pi c \left( 1+z\right)^2 }{H\left( z\right)}d_\mathrm{A}^2\left(z\right) ,
\end{equation}
with angular diameter distance $d_\mathrm{A}\left(z\right)$ and Hubble rate $H\left(z\right)$. For quasi-homogeneous dark energy with sound speed $c_\mathrm{s}=1$ we use the HMF $n\left( M,z\right)$ from~\citet{Tinker08} at an overdensity of $\Delta =200$ with respect to the background density. For cold dark energy with sound speed $c_\mathrm{s}=0$, we use instead the recalibrated HMF of equation~(\ref{eq:dndmcal}) with cosmology-dependent $\delta_{\mathrm{c}}$ (Section~\ref{sec:deltac}) and $\epsilon_{\mathrm{vir}}$ (Section~\ref{sec:epsvir}). It is worth noting here that $\epsilon_{\mathrm{vir}}$ effectively introduces a mass bias between true and observed mass. This is due to the true mass $M$ being effectively shifted to $M\left(1 +\epsilon_\mathrm{vir} \right)$ by the presence of clustering dark energy.

Following~\citet{Lima:2005tt} and assuming a lognormal distribution with variance $\sigma^2_{\ln M}$ we have
\begin{equation}
p\left( M^{\mathrm{ob}} | M ; z\right) = \frac{\exp\left[ -x^2\left( M^{\mathrm{ob}},M,z\right)\right]}{\sqrt{2\sigma^2_{\ln M}(z)}} , \label{eq:pM}
\end{equation}
where $x\left( M^{\mathrm{ob}},M,z\right)$ relates true and observed mass as
\begin{equation}
x\left( M^{\mathrm{ob}},M,z\right) = \frac{\ln M^{\mathrm{ob}}-\ln M_{\mathrm{bias}}(z)-\ln M}{\sqrt{2\sigma^2_{\ln M}(z)}} \, . 
\end{equation}
Inserting equation~(\ref{eq:pM}) into equation~(\ref{eq:Nlm}) and integrating over $\mathrm{d}M^\mathrm{ob}$, we have 
\begin{align}
N_{l,m} = & \frac{\Delta\Omega}{4\pi} \int^{z_{l+1}}_{z_l} \mathrm{d}z \frac{\mathrm{d}V}{\mathrm{d}z} \nonumber \\ & \int^{+\infty}_{0} \mathrm{d}M\, n\left( M,z\right)\left[\mathrm{erfc}\left(x_m\right) - \mathrm{erfc}\left(x_{m+1}\right)\right] \, ,  \label{eq:Nlm2}
\end{align}
with $x_m=x\left( M^{\mathrm{ob}}_{l,m},M,z_l\right)$ and the complementary error function $\mathrm{erfc}\left(x\right)$.\footnote{The complementary error function is defined as \newline $\mathrm{erfc}\left(x\right)=1-\mathrm{erf}\left(x\right)=\frac{2}{\sqrt{\pi}} \int_x^{\infty} e^{-t^2}\mathrm{d}t$.}

As in~\citet{Sartoris10} and~\citet{Sartoris15}, we model the mass-observable relation using the bias between observed and true masses $\ln M_{\mathrm{bias}}$ and  the intrinsic scatter $\sigma^2_{\ln M}$ as
\begin{align}
\ln M_{\mathrm{bias}} \left( z\right) &= B_{\rm M,0} + \alpha\ln \left( 1+z\right) \nonumber \, , \\
\sigma^2_{\ln M} \left( z\right) &= \sigma^2_{\ln M,0} -1 + \left( 1+z\right)^{2\beta} \,, 
\end{align}
with fiducial parameters
\begin{equation}
{\bf p_N} = \left\{  B_{M,0}=0, \,\alpha=0, \, \sigma_{\ln  M,0}=0.4,\, \beta = 0.0\right\} .  \label{eq:nuisance}
\end{equation}
For the quasi-homogeneous dark energy model, this choice of fiducial parameters corresponds to no bias in the mass-observable scaling relation. For cold dark energy, the shift between true and observed mass introduced by $\epsilon_{\mathrm{vir}}$ translates into a mass- and redshift-dependent mass bias. 

\renewcommand{\arraystretch}{1.0}
\setlength{\tabcolsep}{4.5pt}
    \begin{table}
    \centering
        \caption{Survey specifications used for the Fisher matrix forecast of cosmological parameters with cluster number counts (see Section~\ref{sec:Fisher}). Shown are the sky fraction $f_{\mathrm{sky}}$, the threshold mass $M_\mathrm{thr}$, the logarithmic mass binning $\Delta \log \left( M/M_{\odot}\right)$, the maximum survey redshift $z_\mathrm{max}$ and the redshift binning $\Delta z$.}
    \begin{tabular}{| l | c| c| c|  c| c| c| }
    \hline 
    Survey & $f_{\mathrm{sky}}$ (per cent) & $M_{\mathrm{thr}}$ ($M_{\odot}$)  & $\Delta \log \left( M/M_{\odot}\right)$& $z_{\mathrm{max}}$ & $\Delta z$ \\ \hline
     DES & 12 & $1\times10^{14}$ & 0.2 & 1.0 & 0.05 \\
    \hline
    \end{tabular}
    \label{tab:surveys}
    \end{table}

\begin{figure}
\includegraphics[width=0.99\columnwidth]{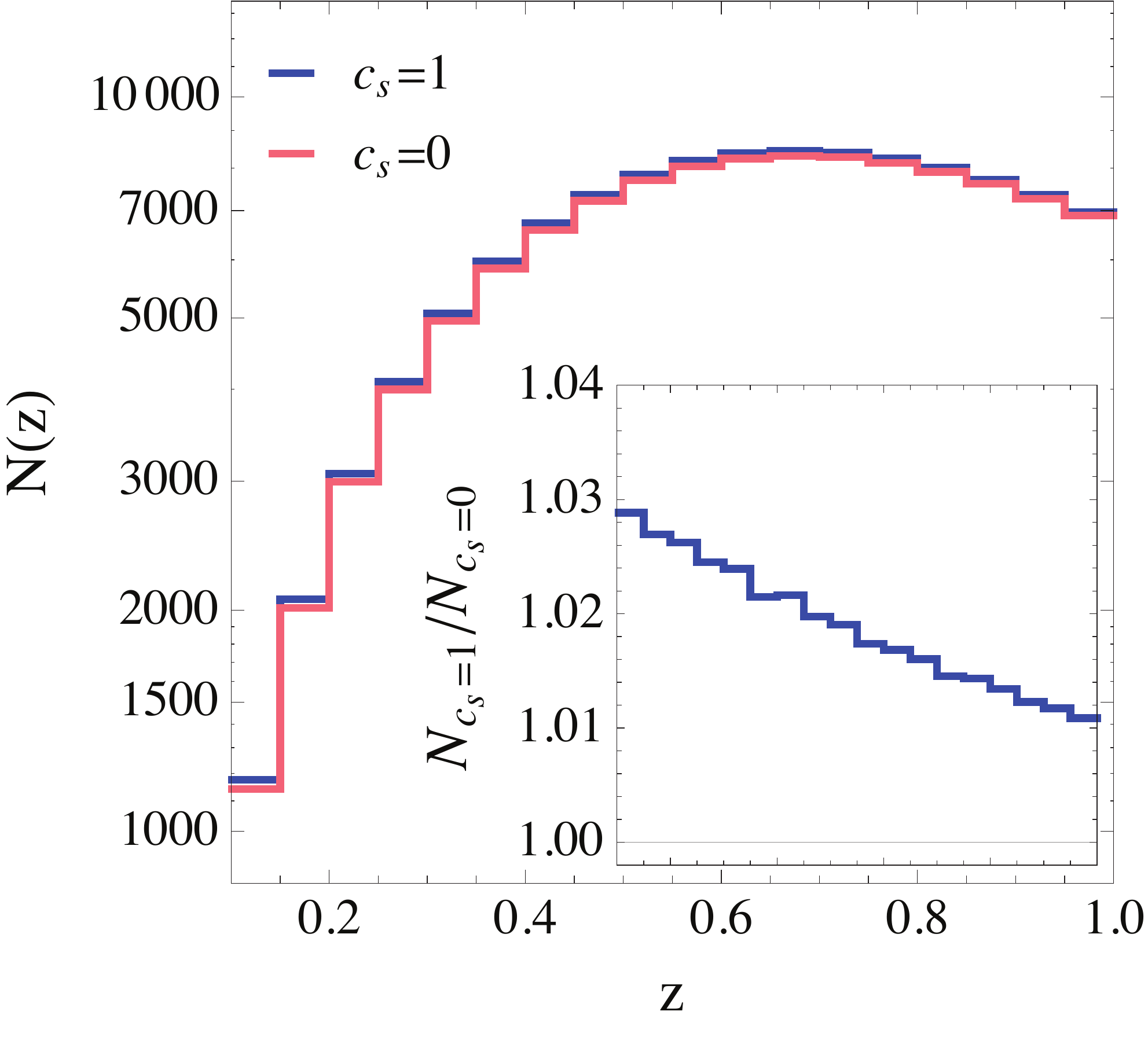}
\caption{Redshift-binned number density of clusters to be detected for $c_\mathrm{s}=1$ (blue line) and $c_\mathrm{s}=0$ (magenta line), and the corresponding ratio of $c_\mathrm{s}=1$ over $c_\mathrm{s}=0$ (inset), for the survey characteristics of DES in Table~\ref{tab:surveys}, and for the fiducial cosmological and nuisance parameter values as stated in the text. Note the slightly higher number of clusters detected in the low-redshift bins for the $c_\mathrm{s}=1$ model, as expected for our fiducial model with $w<-1$ (to be compared with Fig.~\ref{FIG:MFcal} which predicts a higher HMF in this case for $c_\mathrm{s}=1$).}
\label{FIG:Nz}
\end{figure}

We select this form for the mass-observable scaling relation in equation~(\ref{eq:nuisance}), including a mass scatter of $\sigma_{\ln M}=0.4$, to be consistent with the observed mass-richness relation for massive, X-ray-selected clusters at $z<0.5$~\citep{Mantz:2016jyw}. The richness measurements for that relation were drawn from SDSS data using the redMaPPer algorithm~\citep{2014ApJ...785..104R,2015MNRAS.453...38R}, which is a red sequence cluster finder designed to handle an arbitrary photometric galaxy catalogue with excellent photometric redshift performance, completeness and purity, and which is currently in use for Dark Energy Survey (DES) cluster studies.\footnote{http://www.darkenergysurvey.org} We caution, however, that the scatter chosen may be optimistic when applied to optically selected clusters, as in some cases projection effects can boost the measured richness. Also, the redshift evolution of the scatter is not well known and the details will depend on the selection algorithm used. We therefore assume that it remains constant, even though the validity of this assumption may deteriorate at lower masses and higher redshifts, because there the impact of projection effects is expected to increase.

\begin{figure*}
\includegraphics[width=0.65\columnwidth]{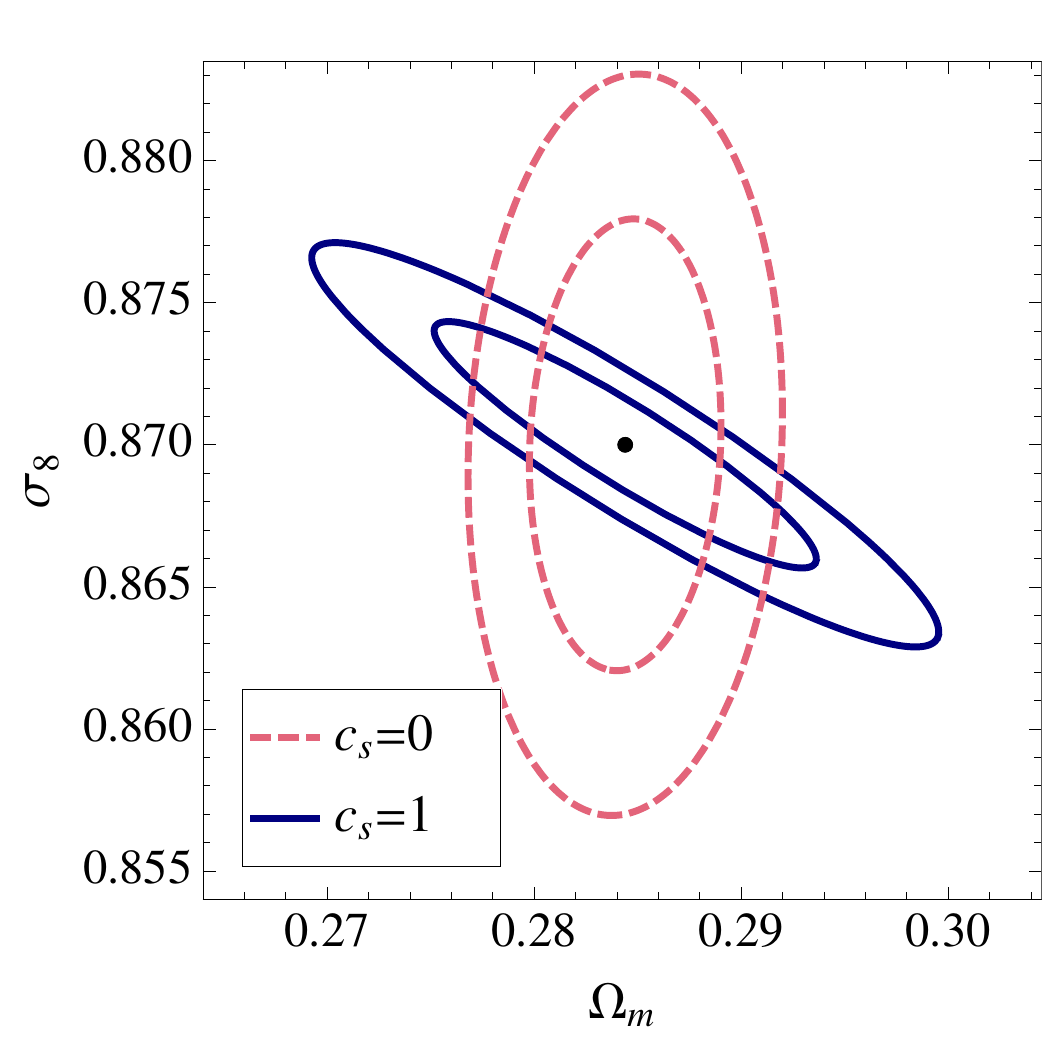}
\hspace{0.09cm}
\includegraphics[width=0.65\columnwidth]{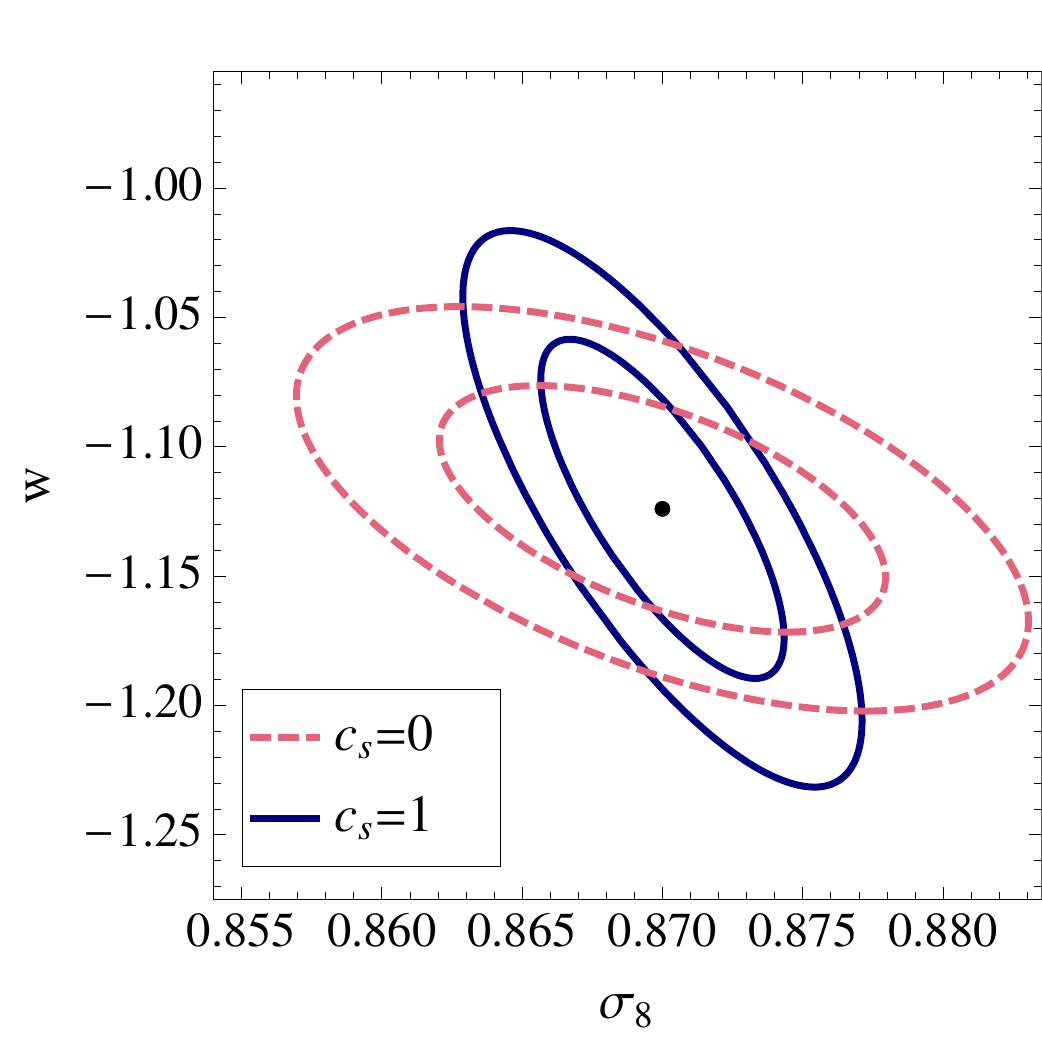}
\hspace{0.09cm}
\includegraphics[width=0.66\columnwidth]{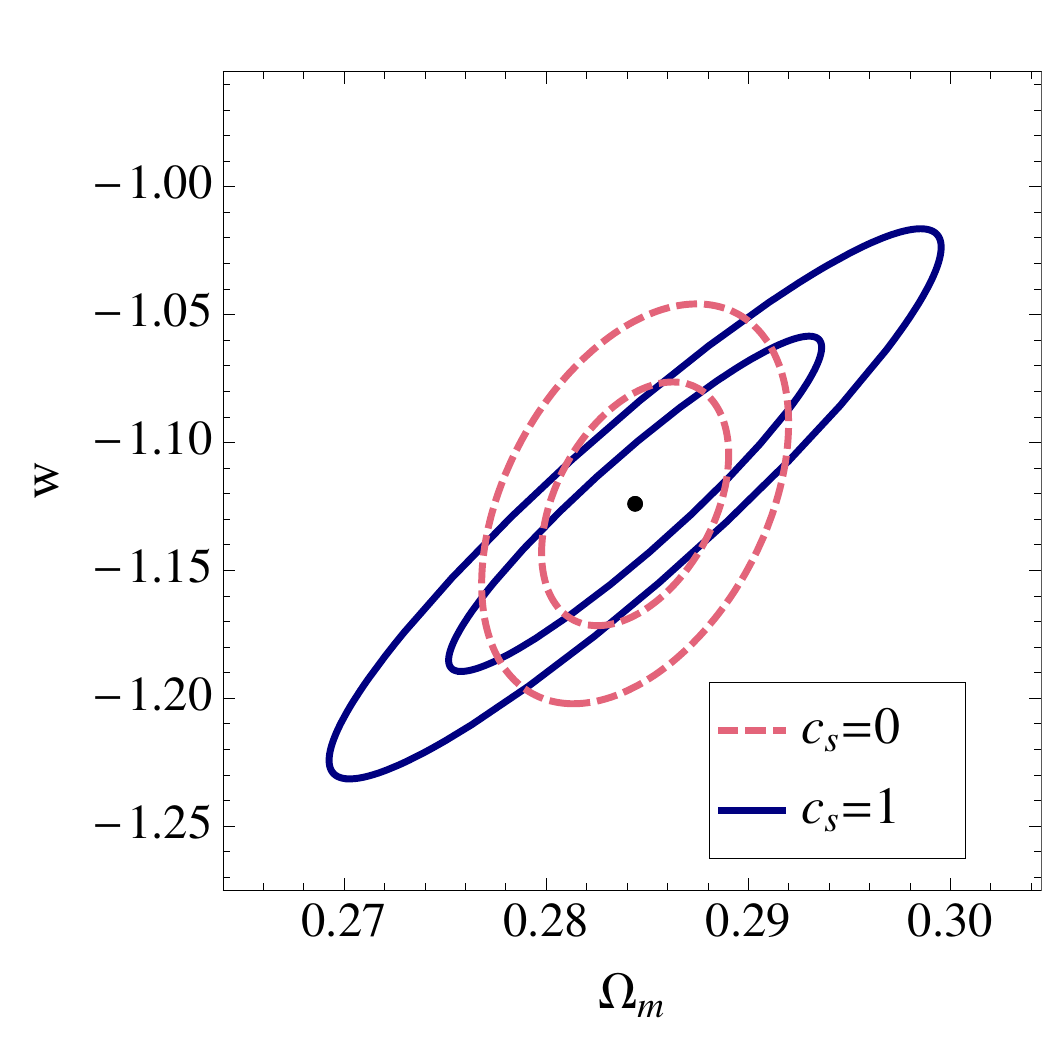}
\caption{Forecasted constraints for DES as described in Section~\ref{sec:Fisher} for $c_\mathrm{s}=1$ (solid, blue contours) and $c_\mathrm{s}=0$ (dashed, magenta contours) at the 68.3 and 95.4 per cent confidence levels. The left-hand panel shows $\sigma_8$ versus $\Omega_\mathrm{m}$; we see that $c_\mathrm{s}=0$ is more constraining in the $\Omega_\mathrm{m}$ direction than for $\sigma_8$ when compared to $c_\mathrm{s}=1$. The middle panel shows $w$ versus $\sigma_8$, where for $c_\mathrm{s}=0$ some constraining power in $w$ is gained, but again at the expense of $\sigma_8$. The right-hand panel shows $w$ versus $\Omega_\mathrm{m}$, now with the constraining power in the $c_\mathrm{s}=0$ case being higher in both $\Omega_\mathrm{m}$ and $w$, as to be expected by the previous two panels. Black dots mark the fiducial model of $\left\{\Omega_\mathrm{m}, \sigma_8,  w \right\}=\left\{ 0.287, 0.87, -1.124 \right\}$.}
\label{FIG:Fisher}
\end{figure*}

 \renewcommand{\arraystretch}{1.0}
\setlength{\tabcolsep}{6pt}
    \begin{table}
    \centering
              \caption{Marginalized 68.3 per cent confidence intervals for DES, for the Fisher matrix forecasts of Section~\ref{sec:Fisher} 
          for both $c_\mathrm{s}=0$ and $c_\mathrm{s}=1$. We also show the FoM in the $\left(\Omega_{\rm m},w \right)$-plane as defined in Section~\ref{sec:Fisher}. The fiducial parameter values are $\left\{\Omega_\mathrm{m}, \sigma_8,  w \right\}=\left\{ 0.287, 0.87, -1.124 \right\}$. The marginalized confidence intervals for the offset in the mass calibration $B_{M,0}$, which was left to vary at the $3\%$ level, are $\Delta B_{M,0}=1.6\%$ for $c_\mathrm{s}=0$ and $\Delta B_{M,0}=0.2\%$ for $c_\mathrm{s}=1$. The tighter constraints on this parameter for $c_\mathrm{s}=1$ are due to the larger number of clusters detected in this case.}
    \begin{tabular}{| c| c| c| c| c| }
    \hline 
    & $\Delta\sigma_8$ (10$^{-3}$)  & $\Delta\Omega_\mathrm{m} $   (10$^{-3}$)  & $\Delta w$ (10$^{-3}$)  & FoM (10$^{3}$)  \\
    \hline
    $c_\mathrm{s}=0$ &  5.2 &  3.1 &  31.4  & \multicolumn{1}{l}{11.4}   \\ 
     $c_\mathrm{s} =1$ &  2.9 &  6.1 &  43.3  & \multicolumn{1}{l}{10.5}  \\ \hline
         \end{tabular}
    \label{tab:Fisher}
    \end{table}

In our Fisher matrix forecast we fix the nuisance parameters $\alpha$, $\sigma_{\ln M,0}$ and $\beta$ to their fiducial values and account for an expected uncertainty in the absolute mass calibration at $\pm3\%$. The cluster mass threshold of detection $M_{\mathrm{thr}}$ is approximated as independent of redshift, although in practice $M_{\mathrm{thr}}$ becomes higher at higher redshifts depending on the survey-specific selection function. For DES, we use a sky coverage of $5000\,$deg$^2$ from $z=0.1$ out to $z=1.0$ with $\Delta z=0.05$ to match the expected photo-$z$ error for masses larger than $M_{\mathrm{thr}}\approx 10^{14}M_{\odot}$~\citep{DES:2016a,DES:2016b}. We bin the mass in steps of $\Delta \log\left( M^{\mathrm{ob}}/M_{\odot}\right)=0.2$ from the threshold mass $M_{\mathrm{thr}}$ up to a maximum of $\log \left( M^{\mathrm{ob}}/M_{\odot}\right)\leq 16$. The survey specifications used for the Fisher matrix forecasts performed in this section are summarized in Table~\ref{tab:surveys}. 
For this analysis, we have implemented our HMF recalibration of Section~\ref{sec:MF} and the Fisher matrix calculation for cluster number counts sketched above into the publicly available {\small \sc CosmoSIS} code~\citep{Zuntz:2014csq}.\footnote{Version 1.4; https://bitbucket.org/joezuntz/cosmosis/wiki} Within {\small \sc CosmoSIS} we ran the provided Fisher matrix sampler together with {\small \sc CAMB},\footnote{Version January 2015.} making use of the existing routine for the calculation of the Tinker and Sheth-Tormen HMFs.\footnote{http://wwwmpa.mpa-garching.mpg.de/$\sim$komatsu/crl/}

Fig.~\ref{FIG:Nz} shows the redshift-binned number density of clusters to be expected for DES in the redshift range $0.1<z<1.0$ for both cold (magenta) and quasi-homogeneous (blue) dark energy. The inset panel shows the corresponding ratio of $c_s=1$ to $c_s=0$, with about $3\%$ more clusters detected in the $z=0.1-0.2$ bin in the $c_s=1$ model.
In Fig.~\ref{FIG:Fisher}, we show the resulting marginalized forecasted constraints for cold (dashed, magenta contours) versus quasi-homogeneous (solid, blue contours) dark energy. We find that constraints on $\Omega_\mathrm{m}$ and $w$ are tighter for cold dark energy. 
This happens though at the expense of those on $\sigma_8$, with less degeneracy between this parameter and $\left(\Omega_\mathrm{m},w\right)$ and the degeneracy direction tilting accordingly. 
For physical intuition, compare the ratio between Sheth--Tormen HMFs for the two sound speeds in the bottom left panel of Fig.~\ref{FIG:MFcal}. This ratio exhibits a strong scale-dependence due to dark energy being clustering, and this dependency is sensitive to both $\Omega_\mathrm{m}$ and $w$, which results in tighter constraints on these parameters.
In general, however, the HMF in the $c_\mathrm{s}=0$ case varies less e.g. when $w$ is varied with respect to $\Lambda$CDM, see for example the top left panel of Fig.~\ref{FIG:MFcal}. This leads to cold dark energy being less constrained at the linear level than quasi-homogeneous dark energy and therefore the uncertainty in $\sigma_8$ being larger, and the degeneracy of $\sigma_8$ with other parameters being smaller.
The differences in marginalized confidence intervals are summed up in Table~\ref{tab:Fisher}, together with a Figure of Merit (FoM), defined similarly to that of the Dark Energy Task Force~\citep[DETF --][]{Albrecht:2006um} as the square root of the determinant of the inverse Fisher matrix, $\mathrm{FoM}=1/\sqrt{|F_i^{-1}|}$, in our case in the $\left(\Omega_m,\,w \right)$-plane. We obtain constraints of the order of $\approx10^{-3}$ for $\Omega_m$ and $\sigma_8$, and $\approx10^{-2}$ for $w$, as well as an FoM $\approx 10^4$, with constraints on $\sigma_8$ being about half and those on $\Omega_\mathrm{m}$ and $w$ being about a factor of 1.5--2 tighter for cold dark energy. 

Note that for {\it Euclid}\footnote{http://www.euclid-ec.org} and Large Synoptic Survey Telescope (LSST)\footnote{https://www.lsst.org/lsst$\_$home.shtml} data, with higher sky coverages and a larger redshift range, as well as better mass calibration, constraints on cosmological parameters are expected to be stronger.
Since both LSST and {\it Euclid} extend to higher redshifts than DES, investigating the impact of cold dark energy with a varying dark energy equation of state should be particularly interesting. Due to the uncertain observable-mass relations at higher redshifts, we postpone such analyses for later studies.

\section{Conclusions}\label{sec:end}
Cold dark energy of sound speed zero presents an interesting phenomenology as compared to standard quasi-homogeneous dark energy with sound speed one. It for example adds on top of the clustering of matter an extra clustering component due to dark energy perturbations that renders the model potentially distinguishable. Within the semi-analytical spherical collapse model, in this paper we derived the non-linear characteristic quantities required for the recalibration of the cluster mass function for cold dark energy. These are the collapse threshold, the virial overdensity and a dark energy mass contribution for cold dark energy. We incorporated these quantities into the HMF, encoding non-linear cosmological model information, to obtain a new cold dark matter plus cold dark energy mass function. 

We use this new recalibrated mass function to constrain cosmological parameters for both quasi-homogeneous dark energy and cold dark energy with current state-of-the-art cluster growth data, together with a combination of standard cosmological data sets. For this combined data set a shift in the best-fitting parameter values of up to $1\sigma$ can be detected, with for example $\sigma_8=0.806\pm0.014$ for sound speed zero and $\sigma_8=0.823\pm0.017$ for sound speed one. These results and the comparison of our recalibrated mass functions for both models makes clear that including further non-linear model information has the potential to distinguish cold dark energy from the standard quasi-homogeneous case, as well as to ease parameter tensions present in the literature (see e.g.~\citet{Kohlinger:2017sxk}).

In order to predict the ability to distinguish cold versus quasi-homogeneous dark energy with upcoming cluster growth data, we made a Fisher matrix forecast of cosmological parameter constraints for the ongoing DES cluster survey, and found $\Omega_m$ and $\sigma_8$ constraints of the order of $10^{-3}$, and of $10^{-2}$ for $w$. Our results also show tighter constraints on $\Omega_\mathrm{m}$ and $w$, increasing them by about 50$\%$ to 100$\%$ for cold dark energy, while the contrary is true for $\sigma_8$, as well as significant differences in the parameter degeneracies between the cold and quasi-homogeneous dark energy models. 

More and better data, as well as combinations with other data, should enhance the differences in the estimated parameters for cold versus quasi-homogeneous dark energy. An interesting direction for further studies would be a more realistic treatment that either allows both the dark energy sound speed and equation of state to vary as free parameters, or employs models for which these parameters naturally evolve with redshift, for example within a Parametrized Post-Friedmann framework~\citep[][]{Hu:2007pj,Hu:2008zd,Fang:2008sn}. Here, using structure formation data in the non-linear regime sensitive to scale-dependent behaviour, which occurs for sound speeds different from zero or one, will prove valuable.

\section*{Acknowledgements}
The computational analysis was performed using the High Performance Computing facility at the University of Copenhagen (HPC@UCPH), and the Gardar supercomputer of the Nordic HPC project. This work was partially supported by the SFB-Transregio TR33 `The Dark Universe' and the DNRF.  DR is supported by a NASA Postdoctoral Program Senior Fellowship at the NASA Ames Research Center, administered by the Universities Space Research Association under contract with NASA. MC acknowledges support from the European Research Council under grant number 647112. We acknowledge support from the U.S. Department of Energy under contract number DE-AC02-76SF00515, and from the National Aeronautics and Space Administration under grant number NNX15AE12G issued through the ROSES 2014 Astrophysics Data Analysis Program. 



\bibliographystyle{mnras}
\bibliography{references} 






\bsp	
\label{lastpage}
\end{document}